\documentclass[11pt]{article}
\usepackage[utf8]{inputenc}
\usepackage{caption}
\usepackage{braket,slashed,bm}
\usepackage{jheppub}
\usepackage{array,multirow}
\usepackage{amsmath} \DeclareMathAlphabet{\mathpzc}{OT1}{pzc}{m}{it}
\usepackage[normalem]{ulem}
\usepackage{calligra}
\usepackage{floatrow}
\DeclareMathAlphabet{\mathpzc}{OT1}{pzc}{m}{it}
\usepackage{mathrsfs}
\usepackage{tikz}

\usepackage{subcaption}
\usepackage{lscape}
\usepackage{hyperref}
\usepackage{graphicx}
\usepackage{booktabs}
\newfloatcommand{capbtabbox}{table}[][\FBwidth]

\graphicspath{{./Figures/}}

\def\beq{\begin{equation}}
\def\eeq{\end{equation}}
\def\bea{\begin{eqnarray}}
\def\eea{\end{eqnarray}}
\def\nn{\nonumber \\}
\def\hyp{\mathsf{y}}

\newcommand{\Lagr}{\mathcal{L}}

\begin{document}
\title{The Geometric Standard Model Effective Field Theory}

\author[a]{Andreas Helset,}

\author[b]{Adam Martin,}

\author[a]{Michael Trott}

\affiliation[a]{Niels Bohr International Academy \& Discovery Center, Niels Bohr Institute, University of Copenhagen,
Blegdamsvej 17, DK-2100, Copenhagen, Denmark}

\affiliation[b]{Department of Physics, University of Notre Dame, Notre Dame, IN, 46556, USA}

\abstract{We develop the geometric formulation of the Standard Model Effective Field Theory (SMEFT).
	Using this approach we derive
all-orders results in the $\sqrt{2 \langle H^\dagger H \rangle}/\Lambda$ expansion
relevant for studies of electroweak precision and Higgs data.}

\maketitle
\section{Introduction}

The Higgs field in the Standard Model (SM) defines a set of field connections of the SM states.
The mass scales of the SM states are dictated by the vaccum expectation value (vev) of the theory, which is defined to be
$\sqrt{2 \langle H^\dagger H\rangle} = \bar{v}_T$.
When the SM is generalized to
the Standard Model Effective Field Theory (SMEFT) \cite{Buchmuller:1985jz,Grzadkowski:2010es},
the Lagrangian contains two characteristic power-counting expansions.
The SMEFT is of interest when physics
beyond the SM is present at scales $\Lambda >\bar{v}_T$.
One of the power-counting expansions in the SMEFT is in the ratio of scales $\bar{v}_T/\Lambda  <1$.
This ratio  defines the nature of the SMEFT operator expansion for measurements with phase space populations
dictated by SM resonances (i.e. near SM poles).
The SMEFT is well-defined and useful when such effects are perturbations to SM predictions.

A second power-counting expansion is present in the SMEFT. This expansion is in $(p^2/\Lambda^2)^{d-4} \lesssim 1$,
where $p^2$ is a kinematic Lorentz invariant.
It is
linked to the novel Lorentz-invariant connections between SM fields, due to higher-dimensional
(and frequently derivative) operators.
This expansion is most relevant when studying measurements with phase space populations
away from the poles of the SM states (when $p^2 \neq m_{SM}^2$), i.e. in
tails of kinematic distributions.

For the SMEFT to be a predictive and meaningful theory, it is
necessary that both of these expansions are under control.\footnote{For this reason, the fact that the number of parameters present in the SMEFT operator basis expansion also grows exponentially on general
grounds \cite{Cardy:1991kr,Henning:2015alf} is a challenge for the SMEFT. We return to this point below.}
In this paper, we develop the geometric approach to the SMEFT. This approach is useful as it
makes the effects of these two distinct power counting expansions
more manifest. Here, we advance this approach by systematically defining
connections that depend on the scalar field coordinates, defining a scalar field space geometry, that is factorized from composite operator forms.
These connections depend on the vev and functionally this is useful as it (largely) factorizes out the $\bar{v}_T/\Lambda$
power counting expansion from the remaining part of the composite operator, and the derivative expansion.
The propagating degrees of freedom, including the Higgs field, then interact on field manifolds, which encode the effects of higher-dimensional operators.
The scalar field space is curved, and the degree of curvature is linked to the size of the
ratio of scales $\bar{v}_T/\Lambda$ \cite{Burgess:2010zq,Alonso:2015fsp,Alonso:2016btr,Alonso:2016oah,Brivio:2017vri,Helset:2018fgq,Corbett:2019cwl}.
This curved field space modifies correlation functions, and the definition of SM Lagrangian parameters such as gauge couplings, mixing angles, and masses.
The flat field space limit of the Lagrangian parameters simplifies to the definitions in the SM as $\bar{v}_T/\Lambda \rightarrow 0$.

In this paper, we also introduce a consistent all-orders general definition of SM Lagrangian parameters (in this expansion) embedded into the SMEFT.
This is possible by taking into account the geometry of the field space defined by the Higgs field.
This definition is limited in form by consistency of the parameter definitions in the SMEFT
$\bar{v}_T/\Lambda$ expansion. These constraints due to self-consistency allow several all-orders results
to be derived for critical experimental observables in electroweak precision and Higgs data, which we also report.

\section{The Geometric SMEFT}

The SMEFT Lagrangian is defined as
\begin{align}
	\Lagr_{\textrm{SMEFT}} &= \Lagr_{\textrm{SM}} + \Lagr^{(d)}, &   \Lagr^{(d)} &= \sum_i \frac{C_i^{(d)}}{\Lambda^{d-4}}\mathcal{Q}_i^{(d)}
	\quad \textrm{ for } d>4.
\end{align}
The particle spectrum has masses $m \sim g_{\rm SM} \sqrt{\langle H^\dagger H \rangle}$, and
includes a $\rm SU(2)_L$ scalar doublet ($H$) with hypercharge
$\hyp_h = 1/2$, distinguishing this theory from the Higgs Effective Field Theory (HEFT) \cite{Feruglio:1992wf,Grinstein:2007iv,Alonso:2012px,Buchalla:2013rka}, where only a singlet scalar is in the
spectrum.\footnote{The direct meaning of this assumption of including a $\rm SU(2)_L$ scalar doublet in the theory,
is that the local operators are {\it analytic} functions of the field $H$. The
analyticity of the operator expansion was reviewed in Ref.~\cite{Brivio:2017vri}.
See also Refs.~\cite{Alonso:2015fsp,Buchalla:2016bse,Chang:2019vez} for some discussion on the HEFT/SMEFT distinction.}

The higher-dimensional operators $\mathcal{Q}_i^{(d)}$ in the SMEFT are labelled with a mass dimension $d$ superscript
and multiply unknown Wilson coefficients $C_i^{(d)}$. The sum over $i$, after non-redundant operators are removed with field redefinitions
of the SM fields, runs over the operators in a particular operator basis.
We use the Warsaw basis \cite{Grzadkowski:2010es} in this paper for $\mathcal{L}^{(6)}$.
The operators defined in Ref.~\cite{Hays:2018zze} are frequently used for $\mathcal{L}^{(8)}$ results, when basis dependent results are quoted.
We frequently absorb powers of $1/\Lambda^2$ into the definition
of the Wilson coefficients for brevity of presentation and use $\tilde{C}^{(6)}_i \equiv C^{(6)}_i \bar{v}_T^2/\Lambda^2$ as a short-hand notation
at times for $\mathcal{L}^{(6)}$ operators.
We generalize this notation to $\mathcal{L}^{(2n)}$ operators, so that
$\tilde{C}^{(2n)}_i \equiv C^{(2n)}_i \bar{v}_T^{2n-4}/\Lambda^{2n-4}$. Our remaining notation is largely consistent with Ref.~\cite{Brivio:2017vri}.

Field space metrics have been studied and developed outside the SMEFT in many works.\footnote{See for example Refs.~\cite{Vilkovisky:1984st,Finn:2019aip}. It is remarkable that
the similar theoretical techniques to those we develop here also enable studies of general relativity as an EFT, see Ref.~\cite{Alonso:2019mok}.}
These techniques are particularly useful in the SMEFT, due to the presence of the Higgs field which takes on a vev.
When this occurs, a tower of high-order field interactions
multiplies a particular composite operator form.
For low $n$-point interactions, the field space metrics defined in Refs.~\cite{Burgess:2010zq,Alonso:2015fsp,Alonso:2016btr,Alonso:2016oah,Helset:2018fgq}
are sufficient to describe this physics.  It has been shown that this approach
can be used to understand what operator forms cannot be removed in operator bases \cite{Burgess:2010zq},
how scalar curvature invariants and the scalar geometry is
related to experiment and the distinction between SMEFT, HEFT and the SM \cite{Alonso:2015fsp,Alonso:2016btr,Alonso:2016oah}, and
how to gauge fix the SMEFT in a manner invariant under background field transformations \cite{Helset:2018fgq}. (See also Ref.~\cite{Misiak:2018gvl}.)
This approach also gives all-orders SMEFT (background field) Ward identities \cite{Corbett:2019cwl}.

The generalization of this approach to arbitrary $n$-point functions is via the decomposition
\bea\label{basicdecomposition}
\Lagr_{\textrm{SMEFT}} = \sum_i f_i(\alpha \cdots) G_i(I,A \cdots),
\eea
where $f_i(\alpha \cdots)$ indicates all explicit Lorentz-index-carrying building blocks of the Lagrangian, while the $G_i$ depend on group indicies $A,I$ for the
(non-spacetime) symmetry groups that act on the scalar fields, and the scalar field coordinates themselves.
By factorizing systematically the dependence on the
scalar field coordinates from the remaining
parts of a composite operator, the expectation value of $G_i(I,A \cdots)$ reduces to a number, and emissions of $h$. This collapses a tower of higher-order interactions into a numerical coefficient
for a composite operator -- when considering matrix elements without propagating $h$ fields.
The $f_i$ are built out of the combinations of fields and derivatives that are outputs of the Hilbert series characterizing and defining a set of higher-dimensional operators,
see Refs.~\cite{Lehman:2015via,Lehman:2015coa,Henning:2015daa,Henning:2015alf,Hays:2018zze}. This introduces a basis dependence into the results.
The Hilbert series generates operator bases with minimal sets of explicit derivatives,
consistent with reductions of operators in an operator basis by the Equation of Motion (EOM). For example, the Warsaw basis for $\mathcal{L}^{(6)}$ is consistent with
the output of a Hilbert series
expansion.\footnote{Such a basis also offers a number of other benefits when calculating in the SMEFT, that are most apparent beyond leading order in the operator expansion;
 see the review \cite{Brivio:2017vri} for more details.} The $f_i$ retain a minimal scalar field coordinate dependence, and vev dependence, through powers of $(D^\mu H)$
 and at higher orders through symmetric derivatives acting on $H$. As these operator forms depend on powers of $\partial_\mu h$ they do not collapse to just a number when
 a scalar expectation value is taken.

 \subsection{Mass eigenstates}
 The field coordinates of the Higgs doublet are put into a convenient form with a common set of generators
 for $\rm SU(2)_L \times U(1)_Y$, by using the real scalar field coordinates $\phi_I = \{\phi_1,\phi_2,\phi_3,\phi_4\}$ introduced with normalization
 \begin{align}
 H(\phi_I) = \frac{1}{\sqrt{2}}\begin{bmatrix} \phi_2+i\phi_1 \\ \phi_4 - i\phi_3\end{bmatrix}, \quad \tilde{H}(\phi_I) = \frac{1}{\sqrt{2}} \begin{bmatrix} \phi_4 + i\phi_3 \\ - \phi_2+i\phi_1\end{bmatrix}.
 \end{align}
$\phi_4$ is expanded around the vacuum expectation value
with the replacement $\phi_4 \rightarrow \phi_4 + \bar{v}_T$.
 The gauge boson field coordinates are similarly unified  into $\mathcal{W}^A = \{W^1,W^2,W^3,B\}$ with $A =\{1,2,3,4\}$.
The corresponding general coupling is defined as $\alpha_A = \{g_2, g_2, g_2, g_1\}$.

We define short-hand notation as in Ref.~\cite{Corbett:2019cwl} for the transformation matrices
that lead to the canonically normalized mass eigenstate fields
 \begin{align*}
 \mathcal{U}^{A}_C &= \sqrt{g}^{AB} U_{BC}, & \quad  \mathcal{V}^{I}_K &= \sqrt{h}^{IJ} V_{JK}.
 \end{align*}
 Here $\sqrt{g}^{AB}$ and $\sqrt{h}^{IJ}$ are square-root metrics, which are understood to be matrix square roots
 of the expectation value --  $\langle \rangle$ -- of the field space connections for the bilinear terms in the SMEFT. These connections are
defined below in Section \ref{scalarnormalization}. The matrices $U,V$ are unitary, and given by
\begin{align*}
	U_{BC} &= \begin{bmatrix}
		\frac{1}{\sqrt{2}} & \frac{1}{\sqrt{2}} & 0 & 0 \\
		\frac{i}{\sqrt{2}} & \frac{-i}{\sqrt{2}} & 0 & 0 \\
		0 & 0 & c_{\overline{\theta}} & s_{\overline{\theta}} \\
		0 & 0 & -s_{\overline{\theta}} & c_{\overline{\theta}}
	\end{bmatrix},& \quad
	V_{JK} &= \begin{bmatrix}
		\frac{-i}{\sqrt{2}} & \frac{i}{\sqrt{2}} & 0 & 0 \\
		\frac{1}{\sqrt{2}} & \frac{1}{\sqrt{2}} & 0 & 0 \\
	0 & 0 & -1 & 0 \\
	0 & 0 & 0 & 1
	\end{bmatrix}.
\end{align*}
Also, $ \sqrt{h}^{IJ} \sqrt{h}_{JK}\equiv \delta^I_K$ and $ \sqrt{g}^{AB} \sqrt{g}_{BC}\equiv \delta^A_C$.
The rotation angles $c_{\overline{\theta}}, s_{\overline{\theta}}$ are functions of $\alpha_A$ and $\langle g^{AB} \rangle$
and are defined geometrically in Section \ref{geometriccouplings}.

The SMEFT weak/mass eigenstate dynamical fields\footnote{The vev $\bar{v}_T$ is subtracted from $\phi_4$ in the equation below involving $\phi^J$.}
and related couplings are then given by \cite{Helset:2018fgq} (see also Refs.~\cite{Alonso:2013hga,Gauld:2015lmb,Dedes:2017zog,Cullen:2019nnr})
\bea\label{massweak}
 \alpha^A = \mathcal{U}^{A}_C \, \beta^C,  \quad  \mathcal{W}^{A,\mu} =  \mathcal{U}^{A}_C \mathcal{A}^{C,\mu},   \quad  \phi^J = \mathcal{V}^{J}_K \, \Phi^K,
\eea
where in the SM limit
\begin{align*}
\alpha^A &= \{g_2 \, g_2, g_2, g_1 \}, & \quad \mathcal{W}^A &= \{W_1,W_2,W_3,B\}, \\
\beta^C &= \left\{\frac{g_2 \,(1-i)}{\sqrt{2}}, \frac{g_2 \,(1+i)}{\sqrt{2}}, \sqrt{g_1^2+ g_2^2}(c_{\bar{\theta}}^2 - s_{\bar{\theta}}^2), \frac{2 \, g_1 \, g_2}{\sqrt{g_1^2+ g_2^2}} \right\}, & \quad \mathcal{A}^{C} &= \left(\mathcal{W}^+,\mathcal{W}^-, \mathcal{Z}, \mathcal{A}\right).
\end{align*}
and $ \phi^J = \{\phi_1,\phi_2,\phi_3,\phi_4 \}, \Phi^K = \{\Phi^-,\Phi^+,\chi, h \}$ for the scalar fields.
All-orders results in the $\bar{v}_T/\Lambda$ expansion can be derived as the relationship between the mass and weak eigenstate
fields is always given by Eqn.~(\ref{massweak}). Remarkably, the remaining field space connections for two- and three-point functions can also be
defined at all-orders in the $\bar{v}_T/\Lambda$ expansion.

\subsection{Classifying field space connections for two-  and three-point functions}\label{sec:classifying}
We first classify the operators contributing to two- and three-point functions.
The arguments used here build on those in
Refs.~\cite{Grzadkowski:2010es,Hays:2018zze}.
Consider a generic three-point function, including the effects of a tower of higher-dimensional operators.
We denote a SM field, defined in the weak eigenstate basis, as $F = \{H, \psi, \mathcal{W}^{\mu \nu} \}$ for the
discussion to follow. Recall the SM EOM for the Higgs field,
\begin{align}
D^2 H_k -\lambda v^2 H_k +2 \lambda (H^\dagger H) H_k + \overline q^j\, Y_u^\dagger\, u (i \sigma_2)_{jk} + \overline d\, Y_d\, q_k +\overline e\, Y_e\,  l_k
&=0 \,,
\label{eomH}
\end{align}
indicating that dependence on $D^2 H_k$ can be removed in a set of operator forms contributing to three-point functions,
in favour of just $H_k$, and higher-point functions. Further, using the Bianchi identity
\bea
D_\mu \mathcal{W}_{\alpha \beta} + D_\alpha \mathcal{W}_{\beta \mu} + D_\beta \mathcal{W}_{\mu \alpha} = 0,
\eea
one can also reduce $D^2 \, \mathcal{W}_{\alpha \beta}$ to EOM-reducible terms and higher-point interactions via
\bea
D^2 \, \mathcal{W}^A_{\alpha \beta} &=& D_{\mu} D_{\nu} g^{\mu \nu} \mathcal{W}^A_{\alpha \beta}, \nn
&=& - D_{\mu} g^{\mu \nu} \left( D_\alpha \mathcal{W}^A_{\beta \nu} + D_\beta \mathcal{W}^A_{\nu \alpha} \right), \nn
&=& - \frac{1}{2} D_{\{\nu, \alpha\}} \mathcal{W}^A_{\beta \nu}  - \frac{1}{2} D_{\{\nu, \beta\}} \mathcal{W}^A_{\nu \alpha}
- \frac{1}{2}\mathcal{W}^A_{\nu \alpha} \mathcal{W}^A_{\beta \nu}
- \frac{1}{2}\mathcal{W}^A_{\nu \beta} \mathcal{W}^A_{\nu \alpha},\nn
&\Rightarrow& \boxed{\rm EOM} \textrm{ and higher-points}
\eea
Here $D_{\{\nu, \alpha\}}$ is the symmetric combination of covariant derivatives. An explicit appearance of $D_{[\mu, \nu]}F$
is reduced to $\mathcal{W}^A_{\mu \nu} F$, where $A$ is dictated by the SM charge of $F$.

Similarly, $D^2 \psi$ can be reduced as
\bea
D^2 \psi = D_\mu D_\nu g^{\mu \nu} \psi = D_\mu D_\nu (\gamma^{\mu}\gamma^{ \nu} + i \sigma^{\mu \nu}) \psi &\Rightarrow& \, \textrm{\boxed{\rm EOM} and higher-points},
\eea
where $\sigma_{\mu\nu}=\frac{i}{2}(\gamma_\mu \gamma_\nu - \gamma_\nu\gamma_\mu)$.
In what follows, when $D^2 F$ appears, it is replaced in terms of EOM terms and higher-point functions for these reasons.
Explicitly reducing operator forms by the EOM, when possible, in favour of other composite operators, has
a key role in these arguments.

Now consider higher-derivative contributions to three-point functions. Explicit appearances of $D^2 F$ are removed
due to the proceeding argument. Further, a general combination of derivatives, acting on three general SM fields $F_{1,2,3}$,
\bea
f(H) (D_\mu F_1) (D_\nu F_2) D_{\{\mu \nu\}} F_3,
\eea
is removable in terms of EOM terms and higher-point functions, using integration by parts:
\begin{align}
	\label{eq:fourderivatives}
	&f(H) (D_\mu F_1) (D_\nu F_2) D_{\{\mu \nu\}} F_3\\
	=&
	-f(H)\left[(D^2 F_1)(D_\nu F_2)+ (D_\mu F_1)(D_\mu D_\nu F_2)
	+ (D_\mu D_\nu F_1)(D_\mu F_2)+ (D_\nu F_1)(D^2 F_2)\right]( D_\nu F_3) \nonumber \\
	&- \left(D_{\mu} f(H)\right)
	\left[(D_\mu F_1)(D_\nu F_2) + (D_\nu F_1)(D_\mu F_2)\right]( D_\nu F_3)\nonumber \\
	\Rightarrow&
	-f(H)\left[(D_\mu F_1)(D_\mu D_\nu F_2)
	+ (D_\mu D_\nu F_1)(D_\mu F_2)\right]( D_\nu F_3)  + \textrm{\boxed{\rm EOM} and higher-points}\nonumber \\
	\Rightarrow&
	-f(H)(D_{[\mu,\nu]} F_1)(D_\mu F_2)( D_\nu F_3)
  + f(H)(D_\mu  F_1)(D_\mu F_2) (D^2 F_3)+ \textrm{ \boxed{\rm EOM} and higher-points} \nonumber \\
	\Rightarrow& \textrm{ \boxed{\rm EOM} and higher-points}. \nonumber
\end{align}
As a result, in general, an operator with four or more derivatives acting on three (possibly different) fields
$F_i$ can be reduced out of three-point amplitudes.

When considering field space connections that can reduce to three-point functions when a vacuum expectation value is taken,
we also use
\begin{align}
\label{eq:derivativesmove}
f(\phi)\, F_1 \, (D_\mu F_2) \, (D_\mu F_3) 	\Rightarrow
(D_\mu f(\phi)) \, (D_\mu F_1) \, F_2 \, F_3 + \frac{1}{2} (D^2 f(\phi)) \, F_1 \, F_2 \, F_3
+ \textrm{ \boxed{\rm EOM}}\, ,
\end{align}
to conventionally move derivative terms onto scalar fields. After reducing the possible field space connections using these arguments
systematically, and integrating by parts, a minimal generalization of field space connections for
$\rm CP$ even electroweak bosonic two- and three-point amplitudes is composed of
\begin{align*}
	h_{IJ}(\phi) (D_\mu \phi)^I (D_\mu \phi)^J,  \quad &g_{AB}(\phi) \mathcal{W}^A_{\mu \nu} \mathcal{W}^{B,\mu \nu},  \quad
	k_{IJ}^A(\phi) (D_\mu \phi)^I (D_\nu \phi)^J \, \mathcal{W}_{A}^{\mu \nu},  \\ \quad &f_{ABC}(\phi)  \mathcal{W}^{A}_{\mu \nu} \mathcal{W}^{B, \nu \rho} \mathcal{W}^{C, \mu}_{\rho},
\end{align*}
and the scalar potential $V(\phi)$.

The minimal set of field space connections involving fermionic field in two- and three-point functions is
\begin{align*}
	Y(\phi) \bar{\psi}_1 \psi_2,  \quad L_{I,A}(\phi) \bar{\psi}_1 \gamma^\mu \sigma_A \psi_2 (D_\mu \phi)^I,  \quad d_A(\phi) \bar{\psi}_1 \sigma^{\mu \nu} \psi_2 \mathcal{W}^A_{\mu \nu},
\end{align*}
where flavour indicies are suppressed.  Here we have defined $\sigma_A = \{\sigma_i, \mathbb{I} \}$, and use this notation below.
The corresponding connections in the case of the gluon field are
\bea
k_{\mathpzc{AB}}(\phi) G^{\mathpzc{A}}_{\mu \nu} G^{\mathpzc{B},\mu \nu},  \quad k_{\mathpzc{ABC}}(\phi) G^{\mathpzc{A}}_{\nu \mu} G^{\mathpzc{B},\rho \nu} G^{\mathpzc{C},\mu \rho},
 \quad c(\phi) \bar{\psi}_1 \sigma^{\mu \nu} T_{\mathpzc{A}} \psi_2 G^{\mathpzc{A}}_{\mu \nu}.
\eea

When considering two- or three-point functions the expectation values of the scalar field connections are
taken with $\langle \rangle$. Although we are focusing our presentation on $\rm CP$ even field space connections,
the case of $\rm CP$ odd connections is analogous and an additional connection can be defined for $g_{AB}$,
$f_{ABC}$, $k_{\mathpzc{AB}}$, and $k_{\mathpzc{ABC}}$. The connections $h_{IJ},g_{AB}$ are symmetric and real, while $f_{ABC}$
and $k_{\mathpzc{ABC}}$ are anti-symmetric.
The $Y(\phi)$, $d_A(\phi)$, and $c(\phi)$ connections are complex. $L_{I,A}$ is real for the SM fields, and complex in general in the
case of the right-handed charged current connection. $k^A_{IJ}$ is antisymmetric in the subscript indicies.

\subsection{Definition of field space connections}\label{scalarnormalization}

The scalar functions include the potential $G_V= V(H^\dagger H)$, with corresponding $f_V \equiv 1$;
\bea
V(\phi) =  \left.-\Lagr_{\rm SMEFT}\right|_{\Lagr(\alpha,\beta\cdots\rightarrow 0)}.
\eea

Going beyond the potential, we define field space connections from the Lagrangian for a series of composite operator forms.
The field space metric for the scalar field bilinear, dependent on the SM field coordinates, is defined via
\bea\label{hijdefn}
h_{IJ}(\phi) = \left.\frac{g^{\mu \nu}}{d} \, \frac{\delta^2 \mathcal{L}_{\rm SMEFT}}{\delta(D_\mu \phi)^I \, \delta (D_\nu \phi)^J} \right|_{\mathcal{L}(\alpha,\beta \cdots) \rightarrow 0}.
\eea
The notation $\mathcal{L}(\alpha,\beta \cdots)$ corresponds to non-trivial Lorentz-index-carrying Lagrangian terms and spin connections, i.e. $\{\mathcal{W}_{\mu \nu}^A, (D^\mu \Phi)^K, \bar{\psi} \sigma^\mu \psi, \bar{\psi} \psi \cdots \}$.
This definition reduces the connection $h_{IJ}$ to a function of $\rm SU(2)_L \times U(1)_Y$ generators, scalar fields coordinates $\phi_i$ and $\bar{v}_T$.

The CP even gauge field scalar manifolds, for the ${\rm SU(2)_L \times U(1)_Y}$ fields interacting with the scalar fields,
are defined as
\bea
g_{AB}(\phi)
=  \left. \frac{-2 \, g^{\mu \nu} \, g^{\sigma \rho}}{d^2}
\, \frac{\delta^2 \mathcal{L}_{\rm SMEFT}}{\delta \mathcal{W}^A_{\mu \sigma} \, \delta \mathcal{W}^B_{\nu \rho}}
 \, \right|_{\mathcal{L}(\alpha,\beta \cdots) \rightarrow 0,{\textrm{CP-even}}},
\eea
and (here $ \mathpzc{A}, \mathpzc{B}$ run over $1 \cdots 8$)
\bea
k_{\mathpzc{AB}}(\phi) =  \left.\frac{-2 \, g^{\mu \nu} \, g^{\sigma \rho}}{d^2} \, \frac{\delta^2 \mathcal{L}_{\rm SMEFT}}{\delta G^{\mathpzc{A}}_{\mu \sigma} \, \delta G^{\mathpzc{B}}_{\nu \rho}} \,\right|_{\mathcal{L}(\alpha,\beta \cdots) \rightarrow 0,{\textrm{CP-even}}}.
\eea
%
We also have
\bea
k_{IJ}^A(\phi) =\left. \frac{g^{\mu\rho}g^{\nu\sigma}}{2d^2} \frac{\delta^3 \mathcal{L}_{\rm SMEFT}}{\delta (D_\mu \phi)^I \delta (D_\nu \phi)^J  \delta \mathcal{W}_{\rho\sigma}^A}\right|_{\mathcal{L}(\alpha,\beta \cdots) \rightarrow 0}
\eea
and
\bea
f_{ABC}(\phi) = \left.\frac{g^{\nu\rho}g^{\sigma\alpha}g^{\beta\mu}}{3!d^3} \frac{\delta^3\mathcal{L}_{\rm SMEFT}}{\delta \mathcal{W}_{\mu\nu}^A \delta \mathcal{W}_{\rho\sigma}^B \delta \mathcal{W}_{\alpha\beta}^C}\right|_{\mathcal{L}(\alpha,\beta\cdots)\rightarrow 0,\textrm{CP-even}},
\nonumber \\
k_{\mathpzc{ABC}}(\phi) = \left.\frac{g^{\nu\rho}g^{\sigma\alpha}g^{\beta\mu}}{3!d^3} \frac{\delta^3\mathcal{L}_{\rm SMEFT}}{\delta G_{\mu\nu}^{\mathpzc{A}} \delta G_{\rho\sigma}^{\mathpzc{B}} \delta G_{\alpha\beta}^{\mathpzc{C}}}\right|_{\mathcal{L}(\alpha,\beta\cdots)\rightarrow 0,\textrm{CP-even}}.
\eea
We also define the fermionic connections
\bea \label{eq:DefYukawa}
Y^{\psi_1}_{pr}(\phi_I) =  \left. \frac{\delta \mathcal{L}_{\rm SMEFT}}{\delta (\bar{\psi}^I_{2,p} \psi_{1,r})} \right|_{\mathcal{L}(\alpha,\beta \cdots) \rightarrow 0},
\quad
L_{J,A}^{\psi,pr} = \left. \frac{\delta^2 \mathcal{L}_{\rm SMEFT}}{\delta (D^\mu \phi)^J \delta (\bar{\psi}_{p} \gamma_\mu \sigma_A \psi_r)}  \right|_{\mathcal{L}(\alpha,\beta \cdots) \rightarrow 0},
\eea
and
\bea
d_A^{\psi_1, pr}(\phi_I) = \left. \frac{\delta^2 \mathcal{L}_{\rm SMEFT}}{\delta (\bar{\psi}_{2,p}^I \sigma_{\mu \nu} \psi_{1,r}) \delta \mathcal{W}_{\mu \nu}^A} \right|_{\mathcal{L}(\alpha,\beta \cdots) \rightarrow 0}.
\eea

\subsection{Hilbert series counting}

The Hilbert series is a compact mathematical tool that uses character orthonormality to count group invariants. As shown in Refs.~\cite{Lehman:2015via,Lehman:2015coa,Henning:2015daa,Henning:2015alf}, it can be adapted to count SMEFT operators up to arbitrary mass dimension while accounting for EOM and integration by parts (IBP) redundancies. The ingredients required are simply the SMEFT field content and each field's representation under the SM gauge groups and 4-d conformal symmetry. The output of the Hilbert series is the number of SMEFT operators with a given mass dimension and field/derivative content. To convert this output into something useful for phenomenology, one must make a choice of how to contract indices and where to apply any derivatives. This choice introduces basis dependence.

The results from the Section \ref{sec:classifying} (combined with similar results from Ref.~\cite{Misiak:2018gvl} for two-point vertices) show that it is possible
to construct a basis where the two- and three-point vertices are particularly simple -- meaning that they are impacted by a minimal set of higher-dimensional operator effects.
Following Eqns.~\eqref{eq:fourderivatives} and \eqref{eq:derivativesmove}, three-point (electroweak) bosonic vertices are captured entirely by operators of the
form $D^2(H^{\dag} H)^n$, $(H^{\dag}H)^n X^2$, $D^2 (H^{\dag}H)^n X$, $(H^{\dag}H)^n X^3$ and $(H^{\dag}H)^n$ ($n$ an integer), with $X_{L/R} = \{W^a \pm \tilde{W}^a,B \pm \tilde{B} ,G \pm \tilde{G}\}$.
The Lorentz group representation is $SO(4) \simeq  \rm SU(2)_L \times SU(2)_R$, so that $X_{L/R}$ are in the $(1,0)$ and $(0,1)$ representations.

Studying the Hilbert series output for this restricted set, we find that the number of invariants in each category approaches a fixed value, and then remains fixed independent of mass dimension: there are $2$ operators of the form $D^2(H^{\dag} H)^n$ for all $n$, 2 operators $(H^{\dag} H)^n\,W^2$, 1 operator $(H^{\dag} H)^n\,WB$, etc.
The fact that the number of operators relevant to the field connections for the two- and three-point vertices saturates can be proven in each case using techniques from Ref.~\cite{Hays:2018zze}. As one example, take $(H^{\dag} H)^n\,W_{L}^2$ and suppress all indices other than Lorentz, in the form $\rm SU(2)_L \otimes SU(2)_R$, and $\rm SU(2)_w$: being bosonic, the $H^n$ and $H^{\dag,n}$ terms must be completely symmetric and therefore in representations $(0,0,\frac n 2)$ of $\rm (SU(2)_L , SU(2)_R, SU(2)_w)$. Their product lies in $(0,0,0\oplus 1\oplus 2\oplus \cdots n)$. $W^2_L$ must also be symmetric, but it is more complicated as $W_L$ contains both Lorentz and $\rm SU(2)_w$ indices (here we use the notation $\rm SU(2)_w$ to avoid a double use of $\rm SU(2)_L$). Keeping all symmetric combinations, we find $(0 \oplus 2,0,0\oplus 2) + (1,0,1)$. Combining the two pieces, the product $(H^{\dag}H)^n W^2_L$ clearly contains two invariants, one where the $(H^{\dag}H)^n$ form a net $\rm SU(2)_w$ singlet, and one where  $(H^{\dag}H)^n$ lie in a quintuplet (spin-2).\footnote{This second possibility requires at least four Higgs fields ($n \ge 2$), and therefore total operator mass dimension $\ge 8$.} Since $B_L$ transforms under Lorentz symmetry alone, there is only one operator of the form $(H^{\dag}H)^n B^2_L$, and the $\rm SU(2)_w$ triplet component of $(H^{\dag}H)^n$ combines with the Lorentz singlet piece of $W_L\, B_L$ to form one operator of the form $(H^{\dag}H)^n W_L\, B_L$. Together, these make up the 4 terms in the $g_{AB}(\phi)\mathcal{W}_{\mu\nu}^A \mathcal{W}^{B,\mu\nu}$ entry of Table \ref{tab:table1} for mass dimension $\ge 8$.\footnote{In addition to the $X^2_L$ operators, there are an identical number of hermitian conjugate terms involving $X_R$. Only one combination of the $X^2_L, X^2_R$ terms are CP conserving.} Similar arguments can be made for the other operator categories in Table \ref{tab:table1}, which are also consistent with the results reported in Ref.~\cite{Henning:2015alf}.

The argument can also be made using on-shell amplitude methods for counting higher-dimensional operators, and
there is clearly a profitable connection between SMEFT geometry and the recent developments using on-shell methods
to study the SMEFT to exploit. See Refs.~\cite{Shadmi:2018xan,Ma:2019gtx,Henning:2019enq,Henning:2019mcv,Durieux:2019eor,Durieux:2019siw,Falkowski:2019zdo,Craig:2019wmo} for recent developments of this form.

Because the number of terms of each operator form for the field connections saturates to a fixed value,
the expressions for the connections for the two- and three-point vertices
at all orders in the $\bar{v}_T/\Lambda$ expansion of the SMEFT can be written compactly and exactly.
This implies that the general exponential nature of the operator basis expansion \cite{Cardy:1991kr,Henning:2015alf}
is more strongly expressed in the growth of higher-point functions and the SMEFT derivative expansion.\footnote{
The very simple form of the resulting field space connections can clearly be examined using
Borel re-summation, once assumptions on perturbativity of the Wilson coefficients are made. This offers the potential
to construct error estimates due to the series truncation on the field space connection. We leave an
exploration of this observation to a future publication.
}
\begin{table}
\begin{tabular}{c|c|c|c|c|c}
\multicolumn{1}{c}{} &
\multicolumn{5}{c}{\textrm Mass Dimension} \\
\hline
\multicolumn{1}{c|}{\textrm Field space connection}
 & 6 & 8 & 10 & 12 & 14 \\
 \hline
 $h_{IJ}(\phi)(D_\mu\phi)^I (D^\mu \phi)^J$ & 2 & 2 & 2 & 2 & 2 \\
 $g_{AB}(\phi)\mathcal{W}_{\mu\nu}^A \mathcal{W}^{B,\mu\nu}$ & 3 & 4 & 4 & 4 & 4 \\
 $k_{IJA}(\phi)(D^\mu \phi)^I(D^\nu\phi)^J \mathcal{W}_{\mu\nu}^A$ & 0 & 3 & 4 & 4 & 4 \\
 $f_{ABC}(\phi) \mathcal{W}_{\mu\nu}^A \mathcal{W}^{B,\nu\rho} \mathcal{W}_{\rho}^{C,\mu} $ & 1 & 2 & 2 & 2 & 2 \\
 $Y^{u}_{pr}(\phi) \bar{Q} u +$ {\textrm h.c.} & $2 \, N_f^2$ & $2 \, N_f^2$ & $2 \, N_f^2$ & $2 \, N_f^2$ & $2 \, N_f^2$ \\
 $Y^{d}_{pr}(\phi) \bar{Q} d +$ {\textrm h.c.} & $2 \, N_f^2$ & $2 \, N_f^2$ & $2 \, N_f^2$ & $2 \, N_f^2$ & $2 \, N_f^2$ \\
 $Y^{e}_{pr}(\phi) \bar{L} e +$ {\textrm h.c.} & $2 \, N_f^2$ & $2 \, N_f^2$ & $2 \, N_f^2$ & $2 \, N_f^2$ & $2 \, N_f^2$ \\
 $d^{e,pr}_A(\phi) \bar{L} \sigma_{\mu \nu} e \mathcal{W}^{\mu \nu}_A +$ {\textrm h.c.} & $4 \, N_f^2$ & $6 \, N_f^2$ & $6 \, N_f^2$ & $6 \, N_f^2$ & $6 \, N_f^2$ \\
 $d^{u,pr}_A(\phi) \bar{Q} \sigma_{\mu \nu} u \mathcal{W}^{\mu \nu}_A +$ {\textrm h.c.} & $4 \, N_f^2$ & $6 \, N_f^2$ & $6 \, N_f^2$ & $6 \, N_f^2$ & $6 \, N_f^2$ \\
 $d^{d,pr}_A(\phi) \bar{Q} \sigma_{\mu \nu} d \mathcal{W}^{\mu \nu}_A +$ {\textrm h.c.} & $4 \, N_f^2$ & $6 \, N_f^2$ & $6 \, N_f^2$ & $6 \, N_f^2$ & $6 \, N_f^2$ \\
$L^{\psi_R}_{pr,A}(\phi) (D^\mu \phi)^J  (\bar{\psi}_{p,R} \gamma_\mu \sigma_A \psi_{r,R}) $  & $N_f^2$ & $N_f^2$ & $N_f^2$ & $N_f^2$ & $N_f^2$ \\
$L^{\psi_L}_{pr,A}(\phi) (D^\mu \phi)^J  (\bar{\psi}_{p,L} \gamma_\mu \sigma_A \psi_{r,L}) $  & $2 \, N_f^2$ & $4 \, N_f^2$ & $4 \, N_f^2$ & $4 \, N_f^2$ & $4 \, N_f^2$
 \end{tabular}
 \caption{\label{tab:table1} Counting of operators contributing to two- and three-point functions from Hilbert series. These results are consistent with Ref.~\cite{Henning:2015alf}.}
\end{table}

\section{Field space connections}
The explicit forms of the field space connections are basis dependent.
In this section we give results in a specific operator basis set, the Warsaw basis at $\mathcal{L}^{(6)}$,
and some operators at $\mathcal{L}^{(8)}$ defined in Ref.~\cite{Hays:2018zze}.

The potential is defined in a power counting expansion as
\bea
V(H^\dagger H) =  \lambda \left(H^\dagger H - \frac{v^2}{2} \right)^2 - C^{(6)}_H (H^\dagger H)^3 - C^{(8)}_{H} (H^\dagger H)^4 \cdots
\eea
The minimum is redefined order by order in the power counting expansion
\bea
\langle H^\dagger H \rangle = \frac{v^2}{2}\left(1+ \frac{3 \, C^{(6)}_H \, v^2}{4 \lambda} + v^4 \frac{9 \, \left(C^{(6)}_H\right)^2 + 4 C_H^{(8)} \lambda}{8 \lambda^2} + \cdots \right) \equiv \frac{\bar{v}_T^2}{2}.
\eea
This generalization of the expectation value simplifies at leading
order in $1/\Lambda^2$ to the vev of the SM. Including the leading $1/\Lambda^2$ correction, the result is that of Ref.~\cite{Alonso:2013hga}, the
$1/\Lambda^4$ correction is as given in Ref.~\cite{Hays:2018zze}, etc. At higher orders in the polynomial expansion of $H^\dagger H$
that results from taking the derivative of the potential, numerical methods must be used to find a minimum due
to the Abel–Ruffini theorem. Note that this also means that expanding out the vev dependence in a formal all-orders result
to a fixed order necessarily requires numerical methods.

The expectation values of the field space connections is also denoted by $\langle \rangle$
and a critical role is played by $\sqrt{h}^{IJ} = \langle h^{IJ} \rangle^{1/2}$, and $\sqrt{g}^{AB}= \langle g^{AB} \rangle^{1/2}$.
The $\sqrt{h}$ and $\sqrt{g}$ depend on $\bar{v}_T$.

\subsection{Scalar bilinear metric:  $h_{IJ}(\phi)$ }

The relevant terms in $\mathcal{L}^{(6,8)}$ for the scalar metric are \cite{Hays:2018zze}
\bea
\mathcal{L}^{(6,8)} &\supseteq& C^{(6)}_{H \Box} (H^\dagger H) \Box (H^\dagger H)+
 C_{HD}^{(6)} (H^\dagger D_\mu H)^\star (H^\dagger D^\mu H) \\
&+& C_{HD}^{(8)} (H^\dagger H)^2 (D_\mu H)^\dagger (D^\mu H) + C_{H,D2}^{(8)}
 (H^\dagger H) (H^\dagger \sigma_a H)  \left[(D_\mu H)^\dagger \, \sigma^a \, (D^\mu H)\right]. \nonumber
\eea
For the Warsaw basis \cite{Misiak:2018gvl}, extended with the $\mathcal{L}^{(8)}$ defined in Ref.~\cite{Hays:2018zze}, $h_{IJ}$ is
\begin{align}\label{hij1}
h_{IJ} = \left[1+ \frac{\phi^4}{4}(C^{(8)}_{HD}- C^{(8)}_{H,D2}) \right] \delta_{IJ}
- 2 C^{(6)}_{H \Box} \phi_I  \phi_J + \frac{\Gamma^I_{A,J} \phi_K \Gamma^K_{A,L} \phi^L}{4} \left(C^{(6)}_{HD} + \phi^2 C_{H,D2}^{(8)} \right).
\end{align}
We note $\delta_{IJ} = \Gamma_{A,K}^I \Gamma_{A,J}^K$ for all $A$ and $\phi^2 = \phi_1^2 + \phi_2^2 + \phi_3^2 + (\phi_4+ \bar{v}_T)^2$.
As we define $h_{IJ}$ as
in Eqn.~(\ref{hijdefn}), the choice in the Warsaw basis to integrate by parts and retain an explicit $ \Box (H^\dagger H)$ derivative
form is notationally awkward. The integration by parts operator identity
\bea
Q_{H \Box}=  (H^\dagger \, i \overleftrightarrow{D}^\mu H ) (H^\dagger \, i \overleftrightarrow{D}_\mu H ) - 4
(H^\dagger D_\mu H)^\star (H^\dagger D^\mu H)
\eea
can be used with the results in the Appendix to write
\bea\label{hij2}
h_{IJ} &=& \left[1+ \frac{\phi^4}{4}(C^{(8)}_{HD}- C^{(8)}_{H,D2}) \right] \delta_{IJ}
+ \frac{\Gamma^I_{A,J} \phi_K \Gamma^K_{A,L} \phi^L}{4} \left(C^{(6)}_{HD} - 4 C^{(6)}_{H \Box}  + \phi^2 C_{H,D2}^{(8)} \right) \nn
&-& 2 (\phi \gamma_{4})_{J} (\gamma_{4} \phi)^I C^{(6)}_{H \Box}.
\eea
Alternatively, one can use field redefinitions, expressed through the EOM operator identity for $\mathcal{L}^{(6)}$ for the Higgs,\footnote{See the Appendix and  Eqn.~(5.3) of Ref.~\cite{Grzadkowski:2010es}.}
to exchange $Q_{H \Box}$ for $H^\dagger H (D^\mu H)^\dagger (D_\mu H)$.
This leads to a redefinition of the Wilson coefficient dependence of the vev
and
\begin{align}
\label{hij3}
h_{IJ} &= \left[1+ \phi^2 C^{(6)}_{H \Box} + \frac{\phi^4}{4}(C^{(8)}_{HD}- C^{(8)}_{H,D2}) \right] \delta_{IJ}
+ \frac{\Gamma^I_{A,J} \phi_K \Gamma^K_{A,L} \phi^L}{4} \left(C^{(6)}_{HD}  + \phi^2 C_{H,D2}^{(8)} \right).
\end{align}
Although the dependence on $C^{(6)}_{H \Box}$ coincides in $\langle h_{IJ} \rangle$ in Eqns.~(\ref{hij1}), (\ref{hij2})
a different dependence on $C^{(6)}_{H \Box}$  is present in $\langle h_{IJ} \rangle$ in Eqn.~(\ref{hij3}).
There is also a redefined vev in this case, and a further correction to the Wilson
coefficient dependence in modified Class five operators in the Warsaw basis, etc. It is important to avoid overinterpreting the
specific, operator basis, and gauge dependent, form of an individual field space connection. Such a quantity,
like a particular Wilson coefficient, in a particular operator basis, is unphysical on its own. (See Appendix \ref{physicalhij} for
more discussion.)
Despite this, a geometric formulation\footnote{Christoffel symbols can be derived from the field space metrics.} of the SMEFT exists in any basis,
and still dictates a consistent relationship between the mass eigenstate field and
the weak eigenstate fields. This relationship also allows all-orders results in the $\bar{v}_T/\Lambda$ expansion
to be derived.

The general form of the scalar metric with $d= 8+ 2n$ dimensional two derivative operators, can be defined
as having the form
\bea
Q_{HD}^{(8+2n)} &=& (H^\dagger H)^{n+2} (D_\mu H)^\dagger (D^\mu H),  \\
 Q_{H,D2}^{(8+2n)} &=&
(H^\dagger H)^{n+1} (H^\dagger \sigma_a H)  (D_\mu H)^\dagger \, \sigma^a \, (D^\mu H), 
\eea
which leads to the result
\bea\label{hij4}
h_{IJ} &=& \left[1+ \phi^2 C^{(6)}_{H \Box} + \sum_{n=0}^\infty \left(\frac{\phi^{2}}{2}\right)^{n+2} \left(C^{(8+ 2n)}_{HD}- C^{(8+2n)}_{H,D2}\right) \right] \delta_{IJ} \nn
       &+& \frac{\Gamma^I_{A,J} \phi_K \Gamma^K_{A,L}\phi^L}{2}  \left(\frac{C^{(6)}_{HD}}{2}  + \sum_{n=0}^\infty \left(\frac{\phi^{2}}{2}\right)^{n+1} \, C_{H,D2}^{(8+ 2n)} \right).
\eea
The scalar field space metric defines a curved field space.

\subsection{Gauge bilinear metric: $g_{AB}(\phi)$ }
The relevant $\mathcal{L}^{(6+ 2n)}$ terms for the Gauge Higgs interactions  are
\bea
 Q_{HB}^{(6+2n)} &=& (H^\dagger H)^{n+1} B^{\mu \nu} \, B_{\mu \nu},  \\
Q_{HW}^{(6+ 2n)} &=&  (H^\dagger H)^{n+1} W_a^{\mu \nu}\, W^a_{\mu \nu}, \\
Q_{HWB}^{(6+ 2n)} &=&  (H^\dagger H)^n (H^\dagger \sigma^a H) W_a^{\mu \nu}\, B_{\mu \nu}, \\
Q_{HW,2}^{(8+ 2n)} &=&  (H^\dagger H)^{n} (H^\dagger \sigma^a H) (H^\dagger \sigma^b H) W_a^{\mu \nu}\, W_{b,\mu \nu}, \\
Q_{HG}^{(6+ 2n)} &=&  (H^\dagger H)^{n+1} G_\mathpzc{A}^{\mu \nu} \, G^\mathpzc{A}_{\mu \nu}.
\eea
The Gauge-Higgs field space metric is given by
\bea
g_{AB}(\phi_I) &=& \left[1 
-4\sum_{n=0}^\infty \left(C_{HW}^{(6+ 2n)}(1- \delta_{A4}) + C_{HB}^{(6+ 2n)} \delta_{A4}\right) \left(\frac{\phi^2}{2}\right)^{n+1} \right]\delta_{AB} \nn
&-&  \sum_{n=0}^\infty C_{HW,2}^{(8+ 2n)} \, \left(\frac{\phi^2}{2}\right)^{n} \left(\phi_I \Gamma_{A,J}^I \phi^J\right) \,  \left(\phi_L \Gamma_{B,K}^L \phi^K\right) (1- \delta_{A4})(1- \delta_{B4}) \nn
&+&\left[\sum_{n=0}^\infty C_{HWB}^{(6+ 2n)} \left(\frac{\phi^2}{2}\right)^n \right]\left[ (\phi_I \Gamma_{A,J}^I \phi^J) \, (1-\delta_{A4})\delta_{B4} + (A\leftrightarrow B) \right],
\eea
and for the gluon fields $G^{\mathpzc{A},\mu} = \sqrt{k}^{\mathpzc{AB}}\, \mathcal{G}_{\mathpzc{B}}^{\mu}$, where
\bea
k_{\mathpzc{AB}}(\phi)= \left(1- 4 \, \sum_{n=0}^\infty C_{HG}^{(6+ 2n)} \, \left(\frac{\phi^2}{2}\right)^n \right) \delta_{\mathpzc{AB}}.
\eea

\subsection{Yukawa couplings: $Y(\phi)$}
The Yukawa interactions of the Higgs field are extended in interpretation in a straightforward manner.
Here the relevant $\mathcal{L}^{(6+ 2n)}$ operators are
\bea
Q^{(6+ 2n)}_{\substack{eH \\ pr}} &=&  (H^\dagger H)^{n+1}(\bar{\ell}_p \, e_r \, H), \\
Q^{(6+ 2n)}_{\substack{uH \\ pr}} &=& (H^\dagger H)^{n+1}  (\bar{q}_p \, u_r \, \tilde{H}), \\
Q^{(6+ 2n)}_{\substack{dH \\ pr}} &=& (H^\dagger H)^{n+1}  (\bar{q}_p \, d_r \, H).
\eea
We define the Yukawa connection in Eqn.~\eqref{eq:DefYukawa},
where
\bea
Y^{e}_{pr}(\phi_I) &=& - H(\phi_I) [Y_e]^\dagger_{pr} + H(\phi_I) \, \sum_{n=0}^\infty C^{(6+ 2n)}_{\substack{eH \\ pr}}  \left(\frac{\phi^2}{2}\right)^{n+1}, \\
Y^{d}_{pr}(\phi_I) &=& - H(\phi_I) [Y_d]^\dagger_{pr} +  H(\phi_I) \,  \sum_{n=0}^\infty C^{(6+ 2n)}_{\substack{dH \\ pr}}  \left(\frac{\phi^2}{2}\right)^{n+1}, \\
Y^{u}_{pr}(\phi_I) &=& - \tilde{H}(\phi_I) [Y_u]^\dagger_{pr} +  \tilde{H}(\phi_I)  \sum_{n=0}^\infty C^{(6+ 2n)}_{\substack{uH \\ pr}}  \left(\frac{\phi^2}{2}\right)^{n+1}.
\eea

\subsection{$(D^\mu\phi)^I \, \bar{\psi} \, \Gamma_\mu \psi$}
The class seven operators in the Warsaw basis, and extended to higher mass dimensions, are of the form
\bea
\mathcal{Q}^{1, (6+2 n)}_{\substack{H \psi\\ pr}} &=& (H^\dagger H)^n H^\dagger \, \overleftrightarrow{i D}^\mu H \bar{\psi}_p \gamma_\mu \psi_r,\nn
\mathcal{Q}^{3, (6+2 n)}_{\substack{H \psi\\ pr}} &=& (H^\dagger H)^n \,H^\dagger \, \overleftrightarrow{i D}_a^\mu H \bar{\psi}_p \gamma_\mu \sigma_a \psi_r,\nn
\mathcal{Q}^{2, (8+2 n)}_{\substack{H \psi\\ pr}} &=& (H^\dagger H)^n (H^\dagger \sigma_a H) \,H^\dagger \, \overleftrightarrow{i D}^\mu H \bar{\psi}_p \gamma_\mu \sigma_a \psi_r,\nn
\mathcal{Q}^{\epsilon, (8+2 n)}_{\substack{H \psi\\ pr}} &=& \epsilon^a_{bc} \, (H^\dagger H)^n \, (H^\dagger \sigma_c H) \,H^\dagger \, \overleftrightarrow{i D}_b^\mu H \bar{\psi}_p \gamma_\mu \sigma_a \psi_r.
\eea
where $\overleftrightarrow{D}^\mu_a = (\sigma_a D^{\mu} - \overleftarrow{D}^{\mu}\,\sigma_a)$.
Connections corresponding to these operators are defined as
\bea
L_{J,A}^{\psi,pr}
&=& - (\phi \, \gamma_{4})_J \delta_{A4}  \sum_{n=0}^\infty C^{1, (6 + 2n)}_{\substack{H \psi\\ pr}} \left(\frac{\phi^2}{2}\right)^n
- (\phi \, \gamma_{A})_J (1- \delta_{A4}) \sum_{n=0}^\infty C^{3, (6 + 2n)}_{\substack{H \psi_L \\ pr}} \left(\frac{\phi^2}{2}\right)^n \nonumber\\
&+&  \frac{1}{2} (\phi \, \gamma_{4})_J (1- \delta_{A4}) \left(\phi_K \Gamma_{A,L}^K \phi^L \right)
\sum_{n=0}^\infty C^{2, (8 + 2n)}_{\substack{H \psi_L \\ pr}} \left(\frac{\phi^2}{2}\right)^n \nn
&+&\frac{\epsilon^A_{BC}}{2}  (\phi \, \gamma_{B})_J \left(\phi_K \Gamma_{C,L}^K \phi^L \right)
\sum_{n=0}^\infty C^{\epsilon, (8 + 2n)}_{\substack{H \psi_L\\ pr}} \left(\frac{\phi^2}{2}\right)^n. 
\eea
Similarly one can define the right-handed charged current
connection
\bea
L^{ud,pr}_J &=& \frac{\delta^2 \mathcal{L}}{\delta (D^\mu \phi)^J \delta (\bar{u}_{p} \gamma_\mu d_r)}
= \frac{\tilde{\phi}_I}{2} (- \Gamma_{4,J}^I + i \gamma_{4,J}^I) \,  \sum_{n=0}^\infty C^{(6 + 2n)}_{\substack{Hud \\ pr}} \left(\frac{\phi^2}{2}\right)^n,
\eea
where $\mathcal{Q}^{(6+ 2n)}_{\substack{Hud\\ pr}}  = (H^\dagger H)^n (\tilde{H} i D^\mu H) \bar{u}_p \gamma_\mu d_r$.

\subsection{$\mathcal{W}_A^{\mu\nu} \, \bar{\psi} \sigma_{\mu \nu} \sigma^A \psi$}
The class six operators in the Warsaw basis, and extended to higher mass dimensions, are of the form
\begin{align*}
	\mathcal{Q}^{(6+2 n)}_{\substack{e W \\ pr}} &= (H^\dagger H)^n \bar{\ell}_p \sigma_{\mu \nu} \sigma^A e_r \mathcal{W}^{\mu \nu}_A H (1- \delta_{A4}),& \quad
	\mathcal{Q}^{(6+2 n)}_{\substack{e B \\ pr}} &= (H^\dagger H)^n \bar{\ell}_p \sigma_{\mu \nu} \sigma^A e_r \mathcal{W}^{\mu \nu}_A H \, \delta_{A4},\nn
	\mathcal{Q}^{(6+2 n)}_{\substack{d W \\ pr}} &= (H^\dagger H)^n \bar{q}_p \sigma_{\mu \nu} \sigma^A d_r \mathcal{W}^{\mu \nu}_A H (1- \delta_{A4}),& \quad
	\mathcal{Q}^{(6+2 n)}_{\substack{d B \\ pr}} &= (H^\dagger H)^n \bar{q}_p \sigma_{\mu \nu} \sigma^A d_r \mathcal{W}^{\mu \nu}_A H \, \delta_{A4},\nn
	\mathcal{Q}^{(6+2 n)}_{\substack{u W \\ pr}} &= (H^\dagger H)^n \bar{q}_p \sigma_{\mu \nu} \sigma^A u_r \mathcal{W}^{\mu \nu}_A \tilde{H} \, (1- \delta_{A4}),& \quad
	\mathcal{Q}^{(6+2 n)}_{\substack{u B \\ pr}} &= (H^\dagger H)^n \bar{q}_p \sigma_{\mu \nu} \sigma^A u_r \mathcal{}W^{\mu \nu}_A \tilde{H} \, \delta_{A4},\nn
\end{align*}
and
\bea
\mathcal{Q}^{(8+2 n)}_{\substack{e W \\ pr}} &=& (H^\dagger H)^n \, (H^\dagger \sigma^A H) \,\bar{\ell}_p \sigma_{\mu \nu} e_r \mathcal{W}^{\mu \nu}_A H (1- \delta_{A4}), \nn
\mathcal{Q}^{(8+2 n)}_{\substack{d W \\ pr}} &=& (H^\dagger H)^n \, (H^\dagger \sigma^A H) \bar{q}_p \sigma_{\mu \nu} d_r \mathcal{W}^{\mu \nu}_A H (1- \delta_{A4}).
\eea
The dipole operator connections are given by
\bea
d_A^{e, pr}
&=& \sum_{n=0}^\infty   \left(\frac{\phi^2}{2}\right)^n \left[\delta_{A4} \,  C^{(6 + 2n)}_{\substack{eB \\ pr}}
+ \sigma_A (1- \delta_{A4})\, C^{(6 + 2n)}_{\substack{eW \\ pr}}
- \left[\phi_K \Gamma_{A,L}^K \phi^L\right] (1- \delta_{A4}) C^{(8 + 2n)}_{\substack{eW2 \\ pr}} \right]H, \nn
d_A^{d, pr}&=&
\sum_{n=0}^\infty   \left(\frac{\phi^2}{2}\right)^n \left[\delta_{A4} \,  C^{(6 + 2n)}_{\substack{dB \\ pr}}
	+ \sigma_A (1- \delta_{A4})\, C^{(6 + 2n)}_{\substack{dW \\ pr}}
- \left[\phi_K \Gamma_{A,L}^K \phi^L\right] (1- \delta_{A4}) C^{(8 + 2n)}_{\substack{dW2 \\ pr}} \right]H , \nn
d_A^{u, pr}&=&
\sum_{n=0}^\infty  \left(\frac{\phi^2}{2}\right)^n \left[\delta_{A4} \,  C^{(6 + 2n)}_{\substack{uB \\ pr}}
	+ \sigma_A (1- \delta_{A4})\, C^{(6 + 2n)}_{\substack{uW \\ pr}}
- \left[\phi_K \Gamma_{A,L}^K \phi^L\right] (1- \delta_{A4}) C^{(8 + 2n)}_{\substack{uW2 \\ pr}} \right]\tilde{H} . \nonumber
\eea
As the Higgs does not carry colour charge, the corresponding connections to gluons are simply 
\bea
c^{u, pr}(\phi)= \tilde{H} \sum_{n=0}^\infty C^{(6 + 2n)}_{\substack{uG \\ pr}} \left(\frac{\phi^2}{2}\right)^n, \quad
c^{d, pr}(\phi) =  H \sum_{n=0}^\infty C^{(6 + 2n)}_{\substack{dG \\ pr}} \left(\frac{\phi^2}{2}\right)^n.
\eea
\subsection{$\mathcal{W}_{A \mu}^{\nu} \, \mathcal{W}_{B \nu}^{\rho} \, \mathcal{W}_{C \rho}^{\mu}$,
$\mathcal{G}_{A \mu}^{\nu} \, \mathcal{G}_{B \nu}^{\rho} \, \mathcal{G}_{C \rho}^{\mu}$}
The relevant operators are
\bea
Q_W^{(6+2n)} &=& \epsilon_{abc}(H^\dagger H)^n W_{\mu\nu}^a W^{\nu\rho,b} W_{\rho}^{\,\,\,\mu, c},  \\
Q_{W2}^{(8+2n)} &=& \epsilon_{abc}(H^\dagger H)^n(H^\dagger \sigma^a H) W_{\mu\nu}^b W^{\nu\rho,c} B_{\rho}^{\,\,\,\mu}, \\
Q_G^{(6+2n)} &=& f_{\mathpzc{ABC}}(H^\dagger H)^n G_{\mu\nu}^{\mathpzc{A}} G^{\nu\rho,\mathpzc{B}} G_{\rho}^{\,\,\,\mu, \mathpzc{C}}.
\eea
The connection for the electroweak fields is given by
\bea
f_{ABC}(\phi) = \epsilon_{ABC} \sum_{n=0}^\infty C^{(6 + 2n)}_{\substack{W}} \left(\frac{\phi^2}{2}\right)^n
- \frac{1}{2} \delta_{A4}\epsilon_{BCD} (\phi_K \Gamma_{D,L}^K \phi^L) \sum_{n=0}^\infty C^{(8 + 2n)}_{\substack{W2}} \left(\frac{\phi^2}{2}\right)^n.
\eea
While in the case of gluon fields it is
\bea
k_{\mathpzc{ABC}}(\phi) = f_{\mathpzc{ABC}} \sum_{n=0}^\infty C^{(6 + 2n)}_{\substack{G}} \left(\frac{\phi^2}{2}\right)^n.
\eea
For both of these connections, there is also a corresponding $\rm CP$ odd connection of a similar form.

\subsection{$(D_\mu \phi)^I \sigma_A (D_\nu \phi)^J \mathcal{W}^A_{\mu \nu}$}
In the Warsaw basis operators of the form $(D_\mu H)^\dagger \sigma_A (D_\nu H) \mathcal{W}^A_{\mu \nu}$
are removed using field redefinitions. This connection is however populated by operator
forms that cannot be removed using field redefinitions,
and a derivative reduction algorithm leading to an operator basis, at higher dimensions.

The form of the connection is given by
\bea
k_{IJ}^A(\phi) &=&  -\frac{1}{2} \gamma_{4,J}^I \delta_{A4} \sum_{n=0}^\infty C^{(8 + 2n)}_{HDHB} \left(\frac{\phi^2}{2}\right)^{n+1}
-\frac{1}{2} \gamma_{A,J}^I (1-\delta_{A4}) \sum_{n=0}^\infty C^{(8 + 2n)}_{HDHW} \left(\frac{\phi^2}{2}\right)^{n+1} \nn
&-&\frac{1}{8}(1-\delta_{A4}) \left[\phi_K  \Gamma_{A,L}^K \phi^L \right] \left[\phi_M  \Gamma_{B,L}^M \phi^N \right]
\gamma_{B,J}^I \sum_{n=0}^\infty C^{(10 + 2n)}_{HDHW,3} \left(\frac{\phi^2}{2}\right)^{n} \\
&+&\frac{1}{4} \epsilon_{ABC} \left[\phi_K  \Gamma_{B,L}^K \phi^L \right] \gamma_{C,J}^I \sum_{n=0}^\infty C^{(8 + 2n)}_{HDHW,2} \left(\frac{\phi^2}{2}\right)^{n}. \nonumber
\eea
Here, the operator forms are defined as
\bea
Q_{HDHB}^{(8+ 2n)} &=& i \, (H^\dagger H)^{n+1}(D_\mu H)^\dagger (D_\nu H) B^{\mu \nu}, \nn
Q_{HDHW}^{(8+ 2n)} &=& i \delta_{ab} (H^\dagger H)^{n+1}(D_\mu H)^\dagger \sigma^a (D_\nu H) W_b^{\mu \nu}, \nn
Q_{HDHW,2}^{(8+ 2n)} &=& i \epsilon_{abc} (H^\dagger H)^{n}(H^\dagger \sigma^a H)(D_\mu H)^\dagger \sigma^b (D_\nu H) W_c^{\mu \nu}, \nn
Q_{HDHW,3}^{(10+ 2n)} &=& i \delta_{ab} \delta_{cd} (H^\dagger H)^{n}(H^\dagger \sigma^a H)(H^\dagger \sigma^c H)
(D_\mu H)^\dagger \sigma^b (D_\nu H) W_d^{\mu \nu}.
\eea

\section{Phenomenology}

\subsection{Higgs mass, and scalar self couplings}
The Higgs mass follows from the potential and is defined as
\bea
\left.    \frac{\delta^2 V(\Phi \cdot \Phi)}{(\delta h)^2} \right|_{\Phi \rightarrow 0} =
2 (\sqrt{h}^{44})^2 \bar{v}_T^2 \left[\frac{\lambda}{2}\left(3 - \frac{v^2}{\bar{v}^2_T}\right)  - \sum_{n=3}^\infty \frac{1}{2^n}
\left(\begin{matrix} 2 n \\ 2 \end{matrix}\right) \tilde{C}_H^{(2n)}
 \right].
\eea
This result follows from $\sqrt{h}^{34}$ vanishing, due to the pseudo-goldstone nature of $\phi^3$.
Similarly the three-, four-, and $m$-point ($m\geq 5$) functions are given by
\bea
\left. -   \frac{\delta^3 V(\Phi \cdot \Phi)}{(\delta h)^3} \right|_{\Phi \rightarrow 0} &=&
 (\sqrt{h}^{44})^3 \, \bar{v}_T \left( - 6 \, \lambda
+ \sum_{n=3}^\infty \frac{1}{2^n}
\left(\begin{matrix} 2 n \\ 3 \end{matrix}\right) \tilde{C}_H^{(2n)}\right),\nn
\left. -   \frac{\delta^4 V(\Phi \cdot \Phi)}{(\delta h)^4} \right|_{\Phi \rightarrow 0} &=&
 (\sqrt{h}^{44})^4 \left( - 6 \, \lambda
+ \sum_{n=3}^\infty \frac{1}{2^n}
\left(\begin{matrix} 2 n \\ 4 \end{matrix}\right) \tilde{C}_H^{(2n)}\right), \nn
\left. -   \frac{\delta^m V(\Phi \cdot \Phi)}{(\delta h)^m} \right|_{\Phi \rightarrow 0} &=&
 (\sqrt{h}^{44})^m \, \sum_{n=3}^\infty \frac{1}{2^n}
\left(\begin{matrix} 2 n \\ m \end{matrix}\right) \tilde{C}_H^{(2n)}.
\eea
\subsection{Fermion masses, and Yukawa couplings}
The fermion masses characterise the intersection of the scalar coordinates with the colour singlet, hypercharge $1/2$ fermion bilinears that can
be constructed out of the SM fermions. The corresponding mass matrices are the expectation value of these field connections
\bea
[M_\psi]_{rp} = \langle (Y^{\psi}_{pr})^\dagger \rangle,
\eea
while the Yukawa interactions are
\bea
[\mathcal{Y}^\psi]_{rp} = \left.\frac{\delta (Y^{\psi}_{pr})^\dagger}{\delta h} \right|_{\phi_i \rightarrow 0}
= \frac{\sqrt{h}^{44}}{\sqrt{2}} \left([Y^\psi]_{rp} - \sum_{n= 3}^\infty \frac{2 n-3}{2^{n-2}} \tilde{C}^{(2n),\star}_{\substack{\psi H \\ pr}} \right).
\eea

\subsection{Geometric definition of gauge couplings}\label{geometriccouplings}

The covariant derivative acting on the scalar fields is
\bea
\left(D^{\mu}\phi\right)^I = \left(\partial^{\mu}\delta_J^I - \frac{1}{2}\mathcal{W}^{A,\mu}\tilde\gamma_{A,J}^I\right)\phi^J,
\eea
with the real generators $\gamma_{A,J}^I$ given in Ref.~\cite{Helset:2018fgq}, and also in the Appendix.
The tilde superscript on $\gamma$ indicates that a coupling dependence has been absorbed into the definition of the generator.
The bilinear terms in the covariant derivative in coupling and field dependence
$g_2 \mathcal{W}^{1,2}_\mu = \bar{g}_2 \mathcal{W}^\pm_
\mu$ etc. remain unchanged due to $\mathcal{L}^{(6)}$ transforming to the
mass eigenstate canonically normalized terms \cite{Alonso:2013hga}.
This corresponds to the invariant $\alpha_A \mathcal{W}^A = \alpha \cdot \mathcal{W}$ being unchanged
by these transformations.
This also holds for corresponding transformations of the QCD coupling and field $g_3 G^\mu = \bar{g}_3 \mathcal{G}^\mu$.
At higher orders in the SMEFT expansion an invariant of this form is also present by construction.
The bar notation is introduced on the couplings to indicate couplings in $\mathcal{L}_{\rm SMEFT}$ that are canonically normalized
as in Ref.~\cite{Alonso:2013hga}. Here this notation also indicates the theory is canonically normalized due to
terms from $\mathcal{L}^{(d>6)}$ that appear in $g^{AB}$.

The geometric definition of the canonically normalized mass eigenstate gauge couplings are
\bea
\bar{g}_2 &=& g_2 \, \sqrt{g}^{11} = g_2 \, \sqrt{g}^{22}, \\
\bar{g}_Z &=& \frac{g_2}{c_{\theta_Z}^2}\left(c_{\bar{\theta}} \sqrt{g}^{33} - s_{\bar{\theta}} \sqrt{g}^{34} \right) = \frac{g_1}{s_{\theta_Z}^2}\left(s_{\bar{\theta}} \sqrt{g}^{44} - c_{\bar{\theta}} \sqrt{g}^{34} \right), \\
\bar{e} &=& g_2\left(s_{\bar{\theta}} \sqrt{g}^{33} + c_{\bar{\theta}} \sqrt{g}^{34}  \right) = g_1\left(c_{\bar{\theta}} \sqrt{g}^{44} + s_{\bar{\theta}} \sqrt{g}^{34}  \right),
\eea
with corresponding mass eigenstate generators listed in the Appendix.  Here we have used the fact that as $\sqrt{g}^{11} = \sqrt{g}^{22}$ due to $\rm SU(2)_L$
gauge invariance, it also follows that $\sqrt{g}^{12} = 0$.
These definitions are geometric and follow directly from
the consistency of the SMEFT description with mass eigenstate fields. These redefinitions hold at all orders in the SMEFT power counting expansion.
Similarly, consistency also dictates the field space geometric definitions of the mixing angles
\begin{align}
s_{\theta_Z}^2 &= \frac{g_1 (\sqrt{g}^{44} s_{\bar{\theta}} - \sqrt{g}^{34} c_{\bar{\theta}})}{g_2 (\sqrt{g}^{33} c_{\bar{\theta}} - \sqrt{g}^{34} s_{\bar{\theta}})+ g_1 (\sqrt{g}^{44} s_{\bar{\theta}} - \sqrt{g}^{34} c_{\bar{\theta}})}, \\
s_{\bar{\theta}}^2 &= \frac{(g_1 \sqrt{g}^{44} - g_2 \sqrt{g}^{34})^2}{g_1^2 [(\sqrt{g}^{34})^2+ (\sqrt{g}^{44})^2]+ g_2^2 [(\sqrt{g}^{33})^2+ (\sqrt{g}^{34})^2]
- 2 g_1 g_2 \sqrt{g}^{34} (\sqrt{g}^{33}+ \sqrt{g}^{44})}.
\end{align}
The gauge boson masses are also defined in a geometric manner as
\bea
\bar{m}_W^2 = \frac{\bar{g}_2^2}{4} \sqrt{h_{11}}^2 \bar{v}_T^2, \quad \quad \bar{m}_Z^2 = \frac{\bar{g}_Z^2}{4}  \sqrt{h_{33}}^2 \bar{v}_T^2 \quad \quad \bar{m}_A^2 = 0.
\eea
To utilize these definitions, and map to a particular operator basis, one must expand out to a fixed order in $\bar{v}_T^2/\Lambda^2$.
Nevertheless, such all-order definitions are of value. The relations hold in any operator basis to define the Lagrangian parameters
incorporating SMEFT corrections in $\bar{v}_T^2/\Lambda^2$ and
clarify the role of these Lagrangian terms in the SMEFT expansion.

 When the covariant derivative acts on fermion fields, the
 Pauli matrices $\sigma_{1,2,3}$ for the $\rm SU(2)_L$ generators\footnote{Defined in the Appendix.}, and the $2 \times 2$ identity matrix $\mathbb{I}$
 for the $\rm U(1)_Y$ generator are used.
 This is a more convenient generator set for chiral spinors.
 The covariant derivative acting on the fermion fields $\psi$, expressed in terms of these quantities,  is
 \bea
 D_\mu \psi &=& \left[\partial_\mu  + i \bar{g}_3 \, \mathcal{G}_{\mathpzc{A}}^\mu \, T^{\mathpzc{A}}  + i \frac{\bar{g}_2}{\sqrt{2}} \left(\mathcal{W}^+ \, T^+ + \mathcal{W}^- \, T^- \right)
 + i \bar{g}_Z \left(T_3 - s_{\theta_Z}^2  Q_\psi \, \right) \mathcal{Z}^\mu + i \, Q_\psi \, \bar{e} \, \mathcal{A}^\mu \right] \psi. \nn
 \eea
Here $Q_\psi = \sigma_3/2 + Y_\psi$ and
the positive sign convention on the covariant derivative is present and the convention
$\sqrt{2} \, \mathcal{W}^\pm = \mathcal{W}^1 \mp i \mathcal{W}^2$ and $\sqrt{2} \, \Phi^\pm = \phi^2 \mp i \phi^1$ is used.
Here $T_3 = \sigma_3/2$ and $2 T^\pm = \sigma_1 \pm i \, \sigma_2$ and $Y_\psi = \{1/6,2/3,-1/3,-1/2,-1 \}$ for $\psi =\{q_L,u_R,d_R,\ell_L,e_R\}$. Note that the $\rm SU(2)_L \times U(1)_Y$
generators of the fermion fields do not need to be the same as those for the scalar and vector fields for
the parameter redefinitions to consistently modify the covariant derivative parameters in the SMEFT.

The covariant derivative acting on the vector fields is defined as
\bea
D_\mu \mathcal{W}^A_\nu = \partial_\mu  \mathcal{W}^A_\nu - \tilde{\epsilon}^{A}_{\, \,BC} \, \mathcal{W}^B_\mu \, \mathcal{W}^C_\nu,
\eea
where the covariant derivative sign convention is consistent with the definition, and also $\mathcal{W}^{A}_{\mu \nu} = \partial_\mu  \mathcal{W}^A_\nu - \partial_\nu  \mathcal{W}^A_\mu - \tilde{\epsilon}^{A}_{\, \,BC} \, \mathcal{W}^B_\mu \, \mathcal{W}^C_\nu$.

\subsection{$\mathcal{W},\mathcal{Z}$ couplings to $\bar{\psi} \psi$}
The mass eigenstate coupling of the $\mathcal{Z}$ and $\mathcal{W}$ to $\bar{\psi} \psi$ are obtained by summing over more than one field space connection.
For couplings to fermion fields of the same chirality, the sum is over $L_{J,A}^{\psi,pr}$ and the modified $\bar{\psi} i \slashed{D} \psi$, that includes the tower of SMEFT corrections in $\mathcal{U}^A_C$.
A compact expression for the mass eigenstate connection is
\bea
- \mathcal{A}^{A,\mu} (\bar{\psi}_{p} \gamma_\mu \bar{\tau}_A \psi_r) \delta_{pr} + \mathcal{A}^{C,\mu}
(\bar{\psi}_{p} \gamma_\mu \sigma_A\psi_r) \langle L_{I,A}^{\psi,pr}\rangle (- {\bm \gamma}^{I}_{C,4}) \bar{v}_T,
\eea
where the fermions are in the weak eigenstate basis. Rotating the fermions to the mass eigenstate basis is straightforward, where the $V_{\rm CKM}$ and $U_{\rm PMNS}$
matrices are introduced as usual. The generators are
\bea
\bar{\tau}_{1,2} = \frac{\bar{g}_2}{\sqrt{2}} \frac{\sigma_1 \pm i \sigma_2}{2}, \quad \bar{\tau}_3 = \bar{g}_Z (T_3 - s_{\theta_Z}^2 Q_\psi), \quad \bar{\tau}_4 = \bar{e} Q_\psi.
\eea
Expanding out to make the couplings explicit, the Lagrangian effective couplings for $\{\mathcal{Z},\mathcal{A},\mathcal{W}^\pm\}$ are
\bea
\langle \mathcal{Z} | \bar{\psi}_{\substack{p}} \psi_{\substack{r}}\rangle &=&
\frac{\bar{g}_Z}{2} \, \bar{\psi}_{\substack{p}} \, \slashed{\epsilon}_{\mathcal{Z}} \, \left[(2 s_{\theta_Z}^2  Q_\psi  - \sigma_3)\delta_{pr}
	+\sigma_3\bar{v}_T \langle L_{3,3}^{\psi,pr}\rangle+ \bar{v}_T \langle L_{3,4}^{\psi,pr} \rangle
 \right] \, \psi_{\substack{r}}, \\
\langle \mathcal{A} | \bar{\psi}_{\substack{p}} \psi_{\substack{r}} \rangle &=&
- \bar{e} \, \bar{\psi}_{\substack{p}} \, \slashed{\epsilon}_{\mathcal{A}} \, Q_\psi \, \delta_{pr}\, \psi_{\substack{r}}, \\
\langle \mathcal{W}_{\pm} | \bar{\psi}_{\substack{p}} \psi_{\substack{r}} \rangle &=& - \frac{\bar{g}_2  }{\sqrt{2}}
\bar{\psi}_{\substack{p}}  (\slashed{\epsilon}_{\mathcal{W}^{\pm}}) \, T^{\pm}
\left[\delta_{pr}-\bar{v}_T \langle L^{\psi,pr}_{1,1}\rangle   \pm  i \bar{v}_T
\langle L^{\psi,pr}_{1,2} \rangle \right]
\, \psi_{\substack{r}}.
\eea
The last expressions simplify due to $\rm SU(2)_L$ gauge invariance. 
Similarly the SMEFT has the right-handed $W^\pm$ couplings to (weak eigenstate) quark fields.
\begin{align*}
\langle \mathcal{W}_{+}^\mu | \bar{u}_{\substack{p}} d_{\substack{r}} \rangle &=   \bar{v}_T \langle L^{ud,pr}_1 \rangle \frac{\bar{g}_2}{\sqrt{2}} \bar{u}_p \,\slashed{\epsilon}_{\mathcal{W}^+} d_r, & \langle \mathcal{W}_{-}^\mu | \bar{d}_{\substack{r}} u_{\substack{p}} \rangle &=   \bar{v}_T \langle L^{ud,pr}_1 \rangle \frac{\bar{g}_2}{\sqrt{2}} \bar{d}_r \,\slashed{\epsilon}_{\mathcal{W}^-} u_p.
\end{align*}

\subsection{Dipole connection of $\mathcal{W},\mathcal{Z}$ to $\bar{\psi} \psi$}
The dipole operators generate a coupling of the $\mathcal{Z}$ and $\mathcal{W}$ to $\bar{\psi} \psi$ that is distinct
from the couplings above, due to the fermion fields being of opposite chirality. Interference between the dipole
connection and the connections in the previous section requires a mass insertion.
The dipole couplings are defined as
\bea
\langle \mathcal{Z} | \bar{u}^{p}_L u^{r}_R \rangle &=&
-2 \bar{g}_Z \bar{u}^p_L \slashed{p}_Z \slashed{\epsilon}_Z u^p_R\,  \left(\langle d_{3}^{u,pr} \rangle \frac{c^2_{\theta_Z}}{g_2}
-\langle d_{4}^{u,pr} \rangle \frac{s^2_{\theta_Z}}{g_1}   \right), \nn
\langle \mathcal{Z} | \bar{d}^{p}_L d^{p}_R \rangle &=&
-2 \bar{g}_Z \bar{d}^p_L \slashed{p}_Z \slashed{\epsilon}_Zd^p_R\, \left(\langle d_{3}^{d,pr} \rangle \frac{c^2_{\theta_Z}}{g_2}
-\langle d_{4}^{d,pr} \rangle \frac{s^2_{\theta_Z}}{g_1}   \right),  \nn
\langle \mathcal{Z} | \bar{e}^{p}_L e^{p}_R \rangle &=&
-2 \bar{g}_Z \bar{e}^p_L \slashed{p}_Z \slashed{\epsilon}_Z e^p_R\,  \left(\langle d_{3}^{e,pr} \rangle \frac{c^2_{\theta_Z}}{g_2}
-\langle d_{4}^{e,pr} \rangle \frac{s^2_{\theta_Z}}{g_1}   \right), 
\eea
and
\bea
\langle \mathcal{W}_{+} | \bar{q}_{\substack{p}} \, d_{\substack{r}} \rangle &=& -\sqrt{2} \frac{\bar{g}_2}{g_2} 
\left( \langle d_{1}^{d,pr} \rangle \,
+i\langle d_{2}^{d,pr} \rangle \, \right)
\bar{u}^p_L \slashed{p}_W \slashed{\epsilon}_W \, d^r_R, \nn
\langle \mathcal{W}_{-} | \bar{q}_{\substack{p}} \, u_{\substack{r}} \rangle &=& -\sqrt{2} \frac{\bar{g}_2}{g_2} 
\left(\langle d_{1}^{u,pr} \rangle \,
-i \langle d_{2}^{u,pr} \rangle \, \right)
\bar{d}^p_L \slashed{p}_W \slashed{\epsilon}_W \, u^r_R, \nn
\langle \mathcal{W}_{+} | \bar{\ell}_{\substack{p}} \, e_{\substack{r}} \rangle &=& -\sqrt{2} \frac{\bar{g}_2}{g_2} 
\left( \langle d_{1}^{e,pr} \rangle \,
+i \langle d_{2}^{e,pr} \rangle \, \right)
\bar{\nu}^p_L \slashed{p}_W \slashed{\epsilon}_W \, e^r_R.
\eea
Here the fermions in the dipole connections are in the weak eigenstate basis and a Hermitian conjugate connection
also exists in each case.
The expectation values of $d_A$ are understood to be the upper (lower) component of an $SU(2)$ doublet
for $d_{1,2}^{e}$, $d_{1,2}^d$, and $d_{3,4}^u$ ($d_{1,2}^u$, $d_{3,4}^e$, and $d_{3,4}^d$).

\subsection{$h \mathcal{A}\mathcal{A}$, $h \mathcal{A}\mathcal{Z}$ couplings}

The effective coupling of {h-$\gamma$-$\gamma$},
including the tower of $\bar{v}_T^2/\Lambda^2$ corrections, is given by
\bea
\langle h| \mathcal{A}(p_1) \mathcal{A}(p_2) \rangle = -\langle h A^{\mu\nu} A_{\mu \nu} \rangle \frac{\sqrt{h}^{44}}{4} \left[
	\langle \frac{\delta g_{33}(\phi)}{\delta \phi_4}\rangle \frac{\overline{e}^2}{g_2^2}  +
 2\langle \frac{\delta g_{34}(\phi)}{\delta \phi_4}\rangle \frac{\overline{e}^2}{g_1 g_2}  +
  \langle \frac{\delta g_{44}(\phi)}{\delta \phi_4}\rangle \frac{\overline{e}^2}{g_1^2}
	\right], \nn
\eea
where $A_{\mu\nu}=\partial_\mu A_\nu - \partial_\nu A_\mu$, and
$\langle h A^{\mu\nu} A_{\mu \nu} \rangle = - 4	(p_1 \! \cdot \! p_2 \epsilon_1 \! \cdot \! \epsilon_2 - p_1 \! \cdot \! \epsilon_2 p_2 \! \cdot \! \epsilon_1 )$
when $\epsilon_1(p_1),\epsilon_2(p_2)$ are the polarization vectors of the external $\gamma$'s.
Similarily the  coupling to {h-$\gamma$-$Z$} is given by
\begin{align}
	\langle h| \mathcal{A}(p_1)& \mathcal{Z}(p_2) \rangle \\ &= -\langle h A^{\mu\nu} Z_{\mu \nu} \rangle \, \frac{\sqrt{h}^{44}}{2}
\bar{e} \, \bar{g}_Z \, \left[
	\langle \frac{\delta g_{33}(\phi)}{\delta \phi_4}\rangle \frac{c_{\theta_Z}^2}{g_2^2}  +
 \langle \frac{\delta g_{34}(\phi)}{\delta \phi_4}\rangle \frac{c_{\theta_Z}^2 - s_{\theta_Z}^2}{g_1 g_2}  -
  \langle \frac{\delta g_{44}(\phi)}{\delta \phi_4}\rangle \frac{s_{\theta_Z}^2}{g_1^2}
	\right], \nonumber 
\end{align}
where $\langle h A^{\mu\nu} Z_{\mu \nu} \rangle = - 2 (p_1 \! \cdot \! p_2 \epsilon_1 \! \cdot \! \epsilon_2 - p_1 \! \cdot \! \epsilon_2 p_2 \! \cdot \! \epsilon_1 )$.

\subsection{$h \mathcal{Z}\mathcal{Z}$, $h \mathcal{W}\mathcal{W}$ couplings}
The off-shell coupling of the Higgs to $\mathcal{Z}\mathcal{Z}$ and $\mathcal{W}\mathcal{W}$
are given by summing over multiple field space connections. One finds
\bea
\langle h| \mathcal{Z}(p_1) \mathcal{Z}(p_2) \rangle &=&
-\frac{\sqrt{h}^{44}}{4} \bar{g}_Z^2 \left[\langle \frac{\delta g_{33}(\phi)}{\delta \phi_4}\rangle \frac{c_{\theta_Z}^4}{g_2^2}  -
2\langle \frac{\delta g_{34}(\phi)}{\delta \phi_4}\rangle \frac{c_{\theta_Z}^2 \, s_{\theta_Z}^2}{g_1 g_2}  +
\langle \frac{\delta g_{44}(\phi)}{\delta \phi_4}\rangle \frac{s_{\theta_Z}^4}{g_1^2}\right]
\langle h \mathcal{Z}_{\mu\nu} \mathcal{Z}^{\mu\nu} \rangle \nn
&+& \sqrt{h}^{44} \frac{\bar{g}_Z^2}{2} \left[\langle\frac{\delta h_{33}(\phi)}{\delta \phi_4}\rangle \left(\frac{\bar{v}_T}{2}\right)^2  + \langle h_{33}(\phi) \rangle \frac{\bar v_T}{2}\, \right]
\langle h \mathcal{Z}_{\mu} \mathcal{Z}^{\mu} \rangle \nn
&+&
 \sqrt{h}^{44} \bar{g}_Z^2 \bar{v}_T \left[\langle k_{34}^3\rangle \frac{c_{\theta_Z}^2}{g_2}-\langle k^4_{34}\rangle \frac{s_{\theta_Z}^2}{g_1} \right]
\langle \partial^\nu h \mathcal{Z}_{\mu} \mathcal{Z}^{\mu \nu} \rangle, 
\eea
and
\bea
\langle h| \mathcal{W}(p_1) \mathcal{W}(p_2) \rangle &=&
-\frac{\sqrt{h}^{44}}{2} \bar{g}_2^2 \left[\langle \frac{\delta g_{11}(\phi)}{\delta \phi_4}\rangle \frac{1}{g_2^2}\right]
\langle h \mathcal{W}_{\mu\nu} \mathcal{W}^{\mu\nu} \rangle \nn
&+& \sqrt{h}^{44} \bar{g}_2^2 \left[\langle\frac{\delta h_{11}(\phi)}{\delta \phi_4}\rangle \left(\frac{\bar{v}_T}{2}\right)^2 + \langle h_{11}(\phi)\rangle\,\frac{\bar v_T}{2} \right]
\langle h \mathcal{W}_{\mu} \mathcal{W}^{\mu} \rangle \nn
&+& 2 \sqrt{h}^{44} \frac{\bar{g}_2^2}{g_2} \frac{\bar{v}_T}{4} \left[i \, \langle k_{42}^1\rangle -
\langle k_{42}^2\rangle  \right] \langle (\partial^\mu h) (\mathcal{W}^+_{\mu\nu} W_-^\nu+ \mathcal{W}^-_{\mu\nu} W_+^\nu)\rangle.
\eea
As these couplings are off-shell, they are not directly observable.

\subsection{$\mathcal{Z} \rightarrow \bar{\psi \psi}$, $\mathcal{W} \rightarrow \bar{\psi \psi}$ partial widths}
A key contribution to the full width of the $\mathcal{Z},\mathcal{W}$ bosons in the SMEFT are the two-body
partial widths that follow from the SMEFT couplings of the $\mathcal{Z},\mathcal{W}$ to fermions of the same chirality.
These results can be defined at all orders in the $\bar{v}_T/\Lambda$ expansion as
\bea
\bar{\Gamma}_{Z \rightarrow \bar{\psi} \psi} = \sum_{\psi}\frac{N_c^\psi}{24 \pi} \sqrt{\bar{m}_Z^2} |g_{\rm eff}^{Z,\psi}|^2 \left(1- \frac{4 \bar{M}_\psi^2}{\bar{m}_Z^2} \right)^{3/2}
\eea
where
\bea
g_{\rm eff}^{Z,\psi} = \frac{\bar{g}_Z}{2} \left[(2 s_{\theta_Z}^2 \, Q_\psi -\sigma_3)\delta_{pr}+ \bar{v}_T \langle L_{3,4}^{\psi,pr} \rangle + \sigma_3
\bar{v}_T \langle L_{3,3}^{\psi,pr} \rangle  \right]
\eea
and  $\psi = \{q_L,u_R,d_R,\ell_L,e_R\}$, while $\sigma_3 = 1$ for $u_L, \nu_L$ and $\sigma_3 = -1$ for $d_L, e_L$.
Similarly one can define
\bea
\bar{\Gamma}_{W \rightarrow \bar{\psi} \psi} = \sum_{\psi}\frac{N_c^\psi}{24 \pi} \sqrt{\bar{m}_W^2} |g_{\rm eff}^{W,\psi}|^2 \left(1- \frac{4 \bar{M}_\psi^2}{\bar{m}_W^2} \right)^{3/2}
\eea
with
\begin{align*}
	g_{\rm eff}^{W,q_L} &= -\frac{\bar{g}_2}{\sqrt{2}} \left[V^{pr}_{\rm CKM} - \bar{v}_T \langle L_{1,1}^{q_L,pr} \rangle \pm i \bar{v}_T \langle L_{1,2}^{q_L,pr} \rangle \right], & \\
	g_{\rm eff}^{W,\ell_L} &= -\frac{\bar{g}_2}{\sqrt{2}} \left[U^{pr,\dagger}_{\rm PMNS} - \bar{v}_T \langle L_{1,1}^{\ell_L,pr} \rangle \pm i \bar{v}_T \langle L_{1,2}^{\ell_L,pr} \rangle \right],
\end{align*}
where the $V_{\rm CKM}$ and $U_{\rm PMNS}$ matrices are implicitly absorbed into $\langle L_{J,A}\rangle$.

\subsection{Higher-point functions}
Field space connections for higher-point functions can also be defined in a straight-forward manner.
However, due to the power-counting expansion in $p^2/\Lambda^2$ and the less trivial kinematic
configurations compared to two- and three-point functions, the number of independent field space
connections for e.g. four-point functions is infinite. This can be seen by noting that the
field space connections can be defined as variations of the Lagrangian with respect to
four fields in the set $\{D_\mu \phi^I, D_{\{\mu,\nu\}} \phi^I, D_{\{\mu,\nu,\rho\}}\phi^I,\cdots\}$,
or analogous sets for $W_{\mu\nu}^A$ or the fermion fields. The higher-derivative terms are the
symmetric combinations of covariant derivatives.

For two- and three-point functions, we used the integration-by-parts relations in Eqns.~\eqref{eq:fourderivatives} and \eqref{eq:derivativesmove}.
This was crucial to make the number field space connections finite and small for two- and three-point functions.
These arguments fail to reduce out higher-derivative field space connections for four-point functions and higher.

The infinite set of field space connections is related to the exponential growth of operators,
and poses a challenge for the practitioners of the SMEFT on general grounds.

\section{Conclusions}
In this paper we have developed the geometric formulation of the SMEFT. This
approach allows all orders results in the $\bar{v}_T/\Lambda$ expansion to be determined. We have developed
and reported several of these results for electroweak precision and Higgs data. 
All-orders expressions are valuable because one can expand directly from the complete result, 
and one need not — potentially laboriously — rederive the result at each order in the $\bar{v}_T/\Lambda$.
These results make manifest
the power, utility and potential of this approach to the SMEFT.

\acknowledgments
We acknowledge support from the Carlsberg Foundation, the Villum Fonden, and the Danish National Research Foundation (DNRF91) through the Discovery center.
The work of AM is partially supported by the National Science Foundation under Grant No. Phy-1230860.
We thank I. Brivio for comments on the manuscript.

\appendix
\section{Generator Algebra}
The Pauli matrices $\sigma_a$, with $a=\{1,2,3 \}$, are given by 
\begin{align}
\sigma_1 &= \left(\begin{matrix} 0 & 1 \\ 1 & 0	\end{matrix} \right), &
\sigma_2 &= \left(\begin{matrix} 0 & -i \\ i & 0	\end{matrix} \right), &
\sigma_3 &= \left(\begin{matrix} 1 & 0 \\ 0 & -1	\end{matrix} \right).
\end{align}
The generators in the real representation are defined as
\begin{align}
	\gamma_{1,J}^{I} =  \begin{bmatrix}
		0 & 0 & 0 & -1 \\
		0 & 0 & -1 & 0 \\
		0 & 1 & 0 & 0 \\
		1 & 0 & 0 & 0
	\end{bmatrix},  \,\,
	\gamma_{2,J}^{I} = \begin{bmatrix}
		0 & 0 & 1 & 0 \\
		0 & 0 & 0 & -1 \\
		-1 & 0 & 0 & 0 \\
		0 & 1 & 0 & 0
	\end{bmatrix}, \,\,
	\gamma_{3,J}^{I} =  \begin{bmatrix}
		0 & -1 & 0 & 0 \\
		1 & 0 & 0 & 0 \\
		0 & 0 & 0 & -1 \\
		0 & 0 & 1 & 0
	\end{bmatrix}, \,\,
	\gamma_{4,J}^{I} = \begin{bmatrix}
		0 & -1 & 0 & 0 \\
		1 & 0 & 0 & 0 \\
		0 & 0 & 0 & 1 \\
		0 & 0 & -1 & 0
	\end{bmatrix}.
\end{align}
We use tilde superscripts when couplings are absorbed in the definition of generators and structure constants,
\begin{align}
	\tilde{\epsilon}^{A}_{\, \,BC} &= g_2 \, \epsilon^{A}_{\, \, BC}, \text{ \, \, with } \tilde{\epsilon}^{1}_{\, \, 23} = +g_2, \text{ \, \, and } \tilde{\epsilon}^{4}_{\, \, BC} = 0,  \nonumber \\
	\tilde{\gamma}_{A,J}^{I} &= \begin{cases} g_2 \, \gamma^{I}_{A,J}, & \text{for } A=1,2,3 \\
		g_1\gamma^{I}_{A,J}, & \text{for } A=4.
					\end{cases}
\end{align}
It is also useful to define a set of matrices
\bea
\Gamma_{A,K}^I = \gamma_{A,J}^{I} \, \gamma_{4,K}^{J}
\eea
where
\begin{align}
	\Gamma_{1,J}^{I} &=  \begin{bmatrix}
		0 & 0 & 1 & 0 \\
		0 & 0 & 0 & -1 \\
		1 & 0 & 0 & 0 \\
		0 & -1 & 0 & 0
	\end{bmatrix}, &
	\Gamma_{2,J}^{I} &= \begin{bmatrix}
		0 & 0 & 0 & 1 \\
		0 & 0 & 1 & 0 \\
		0 & 1 & 0 & 0 \\
		1 & 0 & 0 & 0
	\end{bmatrix}, &
	\Gamma_{3,J}^{I} &=  \begin{bmatrix}
		-1 & 0 & 0 & 0 \\
		0 & -1 & 0 & 0 \\
		0 & 0 & 1 & 0 \\
		0 & 0 & 0 & 1
	\end{bmatrix}, &
	\Gamma_{4,J}^{I} &= - \mathbb{I}_{4 \times 4}.
\end{align}
These matrices have the commutation relations $\left[\gamma_A,\gamma_B \right] = 2 \epsilon^C_{\,\,AB} \, \gamma_C$,
$\left[\gamma_A,\Gamma_B \right] = 2 \epsilon^C_{\,\,AB} \, \Gamma_C$, $\left[\Gamma_A,\Gamma_B \right] = 2 \epsilon^C_{\,\,AB} \, \gamma_C$.
Explicitly the mapping between the generators acting on the field coordinates is $H \rightarrow \sigma_a H$
and $\phi^I \rightarrow - (\Gamma_a)^I_J \, \phi^J$ for $a = \{1,2,3 \}$, while  $H \rightarrow \mathbb{I} \, H$
maps to the real field basis transformation $\phi^I \rightarrow - (\Gamma_4)^I_J \,\phi$. The matrix $\gamma_4$ is used for the hypercharge embedding, and
also plays the role of $i$ in the real representation of the scalar field. $\gamma_4^2 = - \mathbb{I}$ while $i^2 = -1$.
Note that consistent with this the mapping:  $H \rightarrow i \, \sigma_a H$ is related to $\phi^I \rightarrow - (\gamma_a)^I_J \phi^J$,
and $H \rightarrow i \, \mathbb{I} H$ maps to $\phi^I \rightarrow - (\gamma_4)^I_J \, \phi^J$.

An equivalent to complex conjugation is given in the real field basis by
\bea
\gamma_{\star,J}^{I} &=  \begin{bmatrix}
  -1 & 0 & 0 & 0 \\
  0 & 1 & 0 & 0 \\
  0 & 0 & -1 & 0 \\
  0 & 0 & 0 & 1
\end{bmatrix},
\eea
This generator commutes with the remaining generators and $\Gamma_{\star}^2 =  \mathbb{I}$.
Note $\tilde{\phi} = \{\phi_3,\phi_4,-\phi_1,-\phi_2 \}$, and
\bea
H^\dagger \sigma_a H &=& - \frac{1}{2} \phi_I \Gamma^I_{a,J} \phi^J, \\
H^\dagger \, \overleftrightarrow{i D}^\mu H &=& -\phi_I \, \gamma_{4,J}^I(D^\mu \phi)^J  = (D^\mu \phi)_I \, \gamma_{4,J}^I \phi^J,\\
H^\dagger \, \overleftrightarrow{i D}_{a}^\mu H &=& -\phi_I \, \gamma_{a,J}^I(D^\mu \phi)^J = (D^\mu \phi)_I \, \gamma_{a,J}^I \phi^J, \\
2 \tilde{H}^\dagger D^\mu H &=& \tilde{\phi}_I (- \Gamma_{4,J}^I + i \, \gamma_{4,J}^I) \,  (D^\mu \phi)^J.
\eea
Expressing $\tilde{H}^\dagger D^\mu H$ in terms of $\phi$ and $(D^\mu \phi)$ requires the introduction of a singluar matrix,
so the introduction of $\tilde{\phi}$ is preferred.
When considering possible operator forms at higher orders in the SMEFT expansion, it is useful to note that
$\phi_I \Gamma^I_{A,J} \phi^J \neq 0$, while $\phi_I \gamma^I_{a,J} \phi^J = \phi_I \gamma^I_{4,J} \phi^J = 0$.

The transformation of the generators to the mass eigenstate basis is given by
\bea
{\bm \gamma}^{I}_{C,J} = \frac{1}{2} \tilde{\gamma}^{I}_{A,J} \sqrt{g}^{AB} U_{BC}.
\eea
Expanding the results gives the mass eigenstate generators explicitly
\begin{align}
{\bm \gamma}^{I}_{1,J} &= \frac{\bar{g}_2}{2 \sqrt{2}} \left(\gamma^{I}_{1,J}+ i \gamma^{I}_{2,J} \right), & \quad {\bm \gamma}^{I}_{2,J} &= \frac{\bar{g}_2}{2 \sqrt{2}} \left(\gamma^{I}_{1,J}- i \gamma^{I}_{2,J} \right), \\
{\bm \gamma}^{I}_{3,J} &= \frac{\bar{g}_Z}{2} \left(c_{\theta_Z}^2 \gamma^{I}_{3,J} - s_{\theta_Z}^2 \gamma^{I}_{4,J} \right), & \quad {\bm \gamma}^{I}_{4,J} &= \frac{\bar{e}}{2} \left(\gamma^{I}_{3,J} + \gamma^{I}_{4,J} \right).
\end{align}

\section{Physical effects of $\langle h_{IJ} \rangle$}\label{physicalhij}
When $h_{IJ}$ is chosen to have the form
\begin{align}\label{hij4}
h_{IJ} = \left[1+ \frac{\phi^4}{4}(C^{(8)}_{HD}- C^{(8)}_{H,D2}) \right] \delta_{IJ}
- 2 C^{(6)}_{H \Box} \phi_I  \phi_J + \frac{\Gamma^I_{A,J} \phi_K \Gamma^K_{A,L} \phi^L}{4} \left(C^{(6)}_{HD} + \phi^2 C_{H,D2}^{(8)} \right).
\end{align}
then
\begin{align}
	\langle h_{IJ} \rangle = &	 \left[1+ \frac{\bar{v}_T^4}{4}(C^{(8)}_{HD}- C^{(8)}_{H,D2}) \right] \delta_{IJ}
	- 2C^{(6)}_{H \Box} \bar{v}_T^2 \delta_{I,4}  \delta_{J,4} 
\nonumber \\
&+  \frac{\bar{v}_T^2}{2} \,
(\delta_{I,3}  \delta_{J,3} + \delta_{I,4}  \delta_{J,4} ) \left(C^{(6)}_{HD} + \bar{v}_T^2 C_{H,D2}^{(8)} \right).
\end{align}
While if $h_{IJ}$ is chosen to have the form
\begin{align}\label{hij5}
h'_{IJ} &=& \left[1+ \phi^2 C^{(6)}_{H \Box} + \frac{\phi^4}{4}(C^{(8)}_{HD}- C^{(8)}_{H,D2}) \right] \delta_{IJ}
+ \frac{\Gamma^I_{A,J} \phi_K \Gamma^K_{A,L} \phi^L}{4} \left(C^{(6)}_{HD}  + \phi^2 C_{H,D2}^{(8)} \right).
\end{align}
then
\bea
\langle h'_{IJ} \rangle = \left[1+ \bar{v}_T^2 C^{(6)}_{H \Box} + \frac{\bar{v}_T^4}{4}(C^{(8)}_{HD}- C^{(8)}_{H,D2}) \right] \delta_{IJ}
+\frac{\bar{v}_T^2}{2} \,
(\delta_{I,3}  \delta_{J,3} + \delta_{I,4}  \delta_{J,4} ) \left(C^{(6)}_{HD}  + \bar{v}_T^2 C_{H,D2}^{(8)} \right). \nn
\eea
These two cases are related by a field redefinition, expressed through an EOM operator identity at $\mathcal{L}^{(6)}$
\bea
H^\dagger H \Box H^\dagger H &=& 2 (D^\mu H)^\dagger (D_\mu H) H^\dagger H - 2 \lambda v^2 (H^\dagger H) + 4 \lambda (H^\dagger H)^3 \nn
&+& H^\dagger H \left[\overline q^j\, Y_u^\dagger\, (i\sigma_2)_{jk} u + \overline d\, Y_d\, q_k +\overline e\, Y_e\,  l_k  + h.c.\right].
\eea
It is instructive to examine how the difference in the $\Delta \langle h_{IJ} \rangle = \langle h'_{IJ} \rangle - \langle h_{IJ} \rangle$
cancels out of quantities closely related to $S$ matrix elements.  Explicitly
\bea
\Delta \langle h_{IJ} \rangle = \tilde{C}^{(6)}_{H \Box} \left[\delta_{IJ} + 2\delta_{I,4}  \delta_{J,4} \right].
\eea
The modification of $\langle h_{IJ} \rangle$ can be seen to cancel in quantites closely related to $S$-matrix elements, as expected.
For example, one finds
\bea
\Delta [\mathcal{Y}^\psi]_{rp}
= \left[\frac{\Delta \sqrt{h}^{44}}{\sqrt{2}} [Y^\psi]_{rp} - \frac{3}{2 \sqrt{2}} [Y^\psi]_{rp} \, \tilde{C}^{(6)}_{H \Box} \right] = 0.
\eea
for the Yukawa couplings, due to the correlated shift in the $\mathcal{L}^{(6)}$ Yukawa couplings. For the $W$ and $Z$ masses
\bea
\Delta \bar{m}_W^2 = \frac{\bar{g}_2^2}{4} \left[ \Delta \sqrt{h_{11}}^2 \bar{v}_T^2 + \sqrt{h_{11}}^2 \Delta  \bar{v}_T^2 \right] = 0,
\eea
and
\bea
\Delta \bar{m}_Z^2 = \frac{\bar{g}_Z^2}{4} \left[ \Delta \sqrt{h_{33}}^2 \bar{v}_T^2 + \sqrt{h_{33}}^2 \Delta  \bar{v}_T^2 \right] = 0.
\eea
with $\Delta  \bar{v}_T^2 = - \tilde{C}^{(6)}_{H \Box} \, \bar{v}_T^2 $. Conversely, quantites (such as off-shell couplings)
not closely related to $S$-matrix elements,
are not expected to demonstrate an equivalence under field redefinitions, or the transformation of $\langle h_{IJ} \rangle$, and this can be observed
in several off-shell couplings.

\bibliographystyle{JHEP}
\bibliography{bibliography5}

\end{document}